%% file: main-arxiv.tex
\documentclass[11pt,letterpaper,abstract=true]{scrartcl}
\usepackage[margin=1in]{geometry}
\usepackage[T1]{fontenc}
\usepackage{lmodern}
\usepackage{microtype}
\usepackage[UKenglish]{babel}
\usepackage{mathtools}
\usepackage{amssymb,amsthm}
\usepackage{thm-restate}
\usepackage{graphicx}
\usepackage{fontawesome5}
\usepackage{booktabs}
\usepackage{acronym}
\usepackage{xcolor}

\usepackage{bm}

\usepackage{tabularx}
\usepackage{etoolbox}
\usepackage[hidelinks,unicode]{hyperref}
\usepackage[capitalise,noabbrev]{cleveref}
\usepackage[
    backend=biber,
    style=alphabetic,
    minalphanames=3,
    maxalphanames=4,
    maxbibnames=99,
    doi=true,
    giveninits=true
]{biblatex}
\hypersetup{colorlinks=true,linkcolor=blue!40!black,citecolor=blue!40!black,urlcolor=blue!40!black,
    pdftitle={Contiguous Monotone Moldable Job Scheduling},
    pdfsubject={Scheduling, Near Linear Time Algorithm, Approximation}
}

\usepackage{algorithm}
\usepackage{algpseudocode}

\usepackage{standalone}
\usepackage{xspace}
\usepackage{adjustbox}
\usepackage{tikz}
\usetikzlibrary{calc}
\usetikzlibrary{patterns}
\usepackage{subcaption}

\theoremstyle{plain}
\newtheorem{theorem}{Theorem}
\newtheorem{lemma}[theorem]{Lemma}
\newtheorem{observation}[theorem]{Observation}
\theoremstyle{remark}
\newtheorem{remark}[theorem]{Remark}
\newtheorem{claim}[theorem]{Claim}
\theoremstyle{definition}
\newtheorem{definition}[theorem]{Definition}
\theoremstyle{plain}

\newenvironment{claimproof}{%
  \begin{proof}[Proof of Claim]%
}{\end{proof}}

\crefname{property}{Lemma 1.3}{Lemma 1.3}
\Crefname{property}{Lemma 1.3}{Lemma 1.3}
\crefname{observation}{observation}{observations}
\Crefname{observation}{Observation}{Observations}
\crefname{claim}{claim}{claims}
\Crefname{claim}{Claim}{Claims}

\newcommand{\job}{\tilde{\jmath}}
\newcommand{\Aj}{W_{2}}

\newcommand{\qc}{Q(c)}
\newcommand{\qz}{Q(0)}
\newcommand{\opt}{\mathrm{OPT}}
\newcommand{\cpugpu}{CPU/GPU Scheduling\xspace}
\DeclareMathOperator{\load}{load}

\acrodef{cpu}[\textsc{CMMJS}]{Contiguous Monotone Moldable Job Scheduling}
\acrodef{ptas}[\textsc{PTAS}]{Polynomial Time Approximation Scheme}
\acrodef{fptas}[\textsc{FPTAS}]{Fully Polynomial Time Approximation Scheme}
\acrodef{mmckp}[\textsc{MMCKP}]{Multidimensional Multiple-Choice Knapsack Problem}
\acrodef{lpt}[\textsc{LPT}]{Longest-Processing-Time-First}
\acrodef{mckp}[\textsc{MCKP}]{Multiple-Choice Knapsack Problem}

\title{Contiguous Monotone Moldable Job Scheduling}

\author{%
Bennet Edler\thanks{Kiel University, stu235896@mail.uni-kiel.de,\url{https://orcid.org/0009-0002-5067-6781}}
\and
Klaus Jansen\thanks{Kiel University, kj@informatik.uni-kiel.de, \url{https://orcid.org/0000-0001-8358-6796}}
\and
Felix Ohnesorge\thanks{Kiel University, foh@informatik.uni-kiel.de, \url{https://orcid.org/0009-0003-8023-3380}}
}
\date{}
\AfterEndEnvironment{abstract}{%
\par\smallskip
{\small\noindent\textbf{Keywords.} Scheduling, Near Linear Time Algorithm, Approximation.\par
\smallskip\noindent\textbf{Funding.} Deutsche Forschungsgemeinschaft (DFG, German Research Foundation) -- Project number 453769249.}%
}

\begin{document}
\maketitle
\begin{abstract}
    In this paper, we address the Contiguous Monotone Moldable Job Scheduling (CMMJS) problem, where a set of jobs must be scheduled on parallel machines to minimize the overall makespan. Each job is characterized by a processing time function mapping the number of assigned machines to a processing time, such that the corresponding work area function is non-decreasing. Building upon the framework and critical cases identified by Jansen and Ohnesorge (2026), we present an improved analytical approach that yields a tighter approximation ratio of $\frac{2\sqrt{e}}{2\sqrt{e}-1}+\varepsilon \approx 1.4353+\varepsilon$ with a running time of $O(nm\log \frac{1}{\varepsilon})$. We prove that this worst-case analysis is essentially tight by providing a matching lower bound instance. This approximation ratio applies not only to contiguous scheduling but also to the non-contiguous variant of the problem.
    Finally, we demonstrate that our improved bounds extend naturally to heterogeneous environments; seamlessly integrating our results into an existing dual approximation framework for combined CPU/GPU scheduling by Germann~et~al.\ (2024) yields an improved overall approximation ratio of $\frac{2\sqrt{e}}{2\sqrt{e}-1}+\varepsilon$ for the joint problem.
\end{abstract}

\section{Introduction}
In \ac{cpu}, we are given $n \in \mathbb{Z}_{>0}$ jobs and $m\in \mathbb{Z}_{>0}$ machines.
Each job $j\in [n]$ is associated with a function $t_j:[m]\to~\mathbb{Q}_{>0}$ which assigns a number of machines to a processing time.
Furthermore, the work area function $w_j:[m]\to \mathbb{Q}_{>0}$, with \(w_j(k) \coloneqq t_j(k) \cdot k\), defines the work area of a job $j$, when assigned to \(k\) machines.
We assume that $w_j$ is non-decreasing and $t_j$ is non-increasing.
A schedule is given by assigning every job $j$ to a number of machines $\alpha_j\in [m]$ and a starting time $s_j$.
In any feasible schedule, the total number of machines assigned to jobs at any time must not exceed $m$.
A schedule is contiguous if for each job $j$ the machines assigned to it are consecutive, i.e., if $j$ is assigned to machines $m_j,\ldots,m_j+\alpha_j-1$ for some $m_j\in [m-\alpha_j+1]$.
Our goal is to minimize the makespan of the schedule, i.e., $\max_{j\in [n]} s_j + t_j(\alpha_j)$.
We denote the minimal makespan as $\opt$.
For any machine \(i \in [m]\), we say that the load of that machine is the last time where a job has been executed on that machine, i.e. \(\load(i)\coloneqq \max_{j\in [n], i\in [m_j,m_j+\alpha_j)}s_j+t_j(\alpha_j)\).

The general problem of moldable job scheduling is known to be \(\mathrm{NP}\)-hard~\cite{DBLP:journals/siamdm/DuL89a} and without the assumption of monotony, there exists no polynomial-time (\(\frac{3}{2} - \varepsilon\))-approximation, unless $\mathrm{P} = \mathrm{NP}$~\cite{drozdowski1995no32approx,DBLP:journals/scheduling/Johannes06};
Under special assumptions there exist better polynomial-time approximations~\cite{decker200654,jansen2002linear} as well as a pseudo-polynomial (\(\frac{5}{4}+\varepsilon\))-approximation algorithm with running time of \(O(n \log n) \cdot m^{f(\frac{1}{\varepsilon})}\)~\cite{DBLP:conf/esa/JansenR19}, where $f(\cdot)$ is a computable function.
When the number of machines \(m\) is bounded by a polynomial in \(n\), Jansen and Thöle presented a \ac{ptas}~\cite{DBLP:conf/icalp/JansenT08} for the non-contiguous variant as well as a \((\frac{3}{2}+\varepsilon)\)-approximation for the contiguous variant.
Without special restrictions, there exist \(2\)-approximation algorithms for this problem~\cite{turek1992approximate,ludwig1994}.

When assuming that the processing time function \(t_j\) is monotone, the problem is still \(\mathrm{NP}\)-hard~\cite{land2013boundingmoldable}, but the strict \(\frac{3}{2}\) lower bound no longer applies.
Jansen and Land presented a \ac{fptas} for the contiguous variant when ($m > 8\frac{n}{\varepsilon}$)~\cite{DBLP:conf/ipps/JansenL18} and with this generalized the \ac{ptas} for the non-contiguous variant~\cite{DBLP:conf/icalp/JansenT08};
However, the complexity of the \ac{ptas} is roughly \(nm^{2 \cdot 10^{2.4\cdot 10^{17}}}\), even for \(\varepsilon = \frac{1}{2}\).
A similar observation can be made for the pseudo-polynomial-time \((\frac{5}{4}+\varepsilon)\)-approximation algorithm~\cite{DBLP:conf/esa/JansenR19}, where the time complexity resolves to \(\Omega((mn)^{16^{4^{13}}})\) for \(\varepsilon = \frac{1}{4}\).
These large constants have motivated recent work to focus on practically efficient algorithms, i.e., algorithms with a running time of \(O(nm \log(\frac{1}{\varepsilon}))\) or better.

Mounié, Rapine, and Trystram presented a \((\frac{3}{2} + \varepsilon)\)-approximation algorithm~\cite{DBLP:journals/siamcomp/MounieRT07}.
The complexity of this algorithm was later improved by Jansen and Land~\cite{DBLP:conf/ipps/JansenL18} and Grage et al.~\cite{DBLP:conf/europar/GrageJO23}.
While not explicitly stated in the original work, theses algorithms can be adapted to the contiguous case as well~\cite{ohnesorge}.
Wu, Zhang, and Chen~\cite{DBLP:journals/eor/WuZC23} eliminated the \(\varepsilon\)-factor in the approximation ratio, achieving a \(\frac{3}{2}\)-approximation algorithm.
Just recently, Jansen and Ohnesorge~\cite{ohnesorge} presented an algorithm with a worst-case ratio of roughly \(1.4593 + \varepsilon\) for the contiguous variant.
\begin{table}[t]
    \caption{Practically efficient algorithms for Monotone Moldable Job Scheduling. The $\tilde{O}$ notation hides logarithmic factors in $n$, $m$, and $1/\varepsilon$. Algorithms marked with (*) were originally designed for the non-contiguous case but can be adapted to the contiguous case~\cite{ohnesorge}.}
    \label{table:ratios}
    \centering
    \begin{tabular}{llll}
        \textbf{Ratio}              & \textbf{Runtime}             & \textbf{Contiguous?} & \textbf{Author}                                                          \\
        \hline
        $\frac{3}{2} + \varepsilon$ & $O(nm \log(1/\varepsilon))$  & Yes*                 & Mounié, Rapine, Trystram (2007) \cite{DBLP:journals/siamcomp/MounieRT07} \\

        $\frac{3}{2} + \varepsilon$ & $\tilde{O}(n/\varepsilon^3)$ & Yes*                 & Jansen, Land (2018) \cite{DBLP:conf/ipps/JansenL18}                      \\

        $\frac{3}{2} + \varepsilon$ & $\tilde{O}(n/\varepsilon)$   & Yes*                 & Grage, Jansen, Ohnesorge (2023) \cite{DBLP:conf/europar/GrageJO23}       \\

        $\frac{3}{2}$               & $\tilde{O}(nm)$              & No                   & Wu, Zhang, Chen (2023) \cite{DBLP:journals/eor/WuZC23}                   \\

        $1.4593 + \varepsilon$      & $O(nm \log(1/\varepsilon))$  & Yes                  & Jansen, Ohnesorge (2026) \cite{ohnesorge}                                \\

        $1.4353 + \varepsilon$      & $O(nm \log(1/\varepsilon))$  & Yes                  & \textbf{This Work}                                                       \\
        \hline
    \end{tabular}%
\end{table}
Note that the running time \(O(nm)\) is technically only pseudo-polynomial, but since Jansen and Land~\cite{DBLP:conf/ipps/JansenL18} developed a \ac{fptas} when the number of machines is large ($m>8\frac{n}{\varepsilon}$), we can assume that $m$ is polynomially in $n$ to make the runtime polynomial.

The perhaps most useful observation for designing algorithms for this monotone variant is that we can assign fewer machines to a job without increasing the work area.
This has been used by Jansen and Land~\cite{DBLP:conf/ipps/JansenL18} to round jobs that are assigned to many machines.
\begin{definition}
    For any $j \in [n]$ and $\alpha' < \alpha \leq m$, the monotony implies:
    \begin{align}
        t_j(\alpha') \leq \frac{w_j(\alpha)}{\alpha'} = t_j(\alpha) \frac{\alpha}{\alpha'}.
    \end{align}
\end{definition}
For example, when we assign a job to one instead of two machines, the new processing time will double at most.
Similar to previous work~\cite{DBLP:conf/europar/GrageJO23,ohnesorge,DBLP:journals/siamcomp/MounieRT07}, we will heavily rely on this property to prove our main result:

\begin{restatable}{theorem}{FullTheoremCmd}
    \label{thm:full}
    Let $\varepsilon > 0$. For \acl{cpu}, there exists an algorithm with an approximation ratio of $\frac{2\sqrt{e}}{2\sqrt{e}-1}+\varepsilon$ and runtime of $O(nm\log \frac1\varepsilon)$.
\end{restatable}

It is worth noting that our algorithm, while designed for the contiguous case, achieves the exact same approximation ratio for the non-contiguous variant. This property is not trivially given, since the optimal makespan for a non-contiguous schedule can be strictly smaller than the optimal makespan for a contiguous one. However, our analysis--similar to the approaches used for other recent contiguous algorithms~\cite{DBLP:conf/europar/GrageJO23,ohnesorge}--uses the maximal total work area to bound the resulting makespan. Consequently, our approximation ratio of $\frac{2\sqrt{e}}{2\sqrt{e}-1}+\varepsilon$ applies to the non-contiguous case as well.

To complement this upper bound, we also provide a tight lower bound example in \Cref{app:instance} demonstrating that the worst-case analysis of the algorithm is tight:

\begin{restatable}{lemma}{LowerBoundLemmaCmd}
    \label{lem:lower_bound}
    For any \(\varepsilon>0\), there exists an instance \(I\) for which \Cref{alg:jansen_ohnesorge} returns a schedule with makespan \(\mathrm{ALG}(I)\) and \(\frac{\mathrm{ALG(I)}}{\mathrm{OPT}(I)} \geq\frac{2\sqrt{e}}{2\sqrt{e}-1}-\varepsilon\).
\end{restatable}

Furthermore, we demonstrate that our improved bounds for contiguous monotone moldable job scheduling extend naturally to heterogeneous environments.
Specifically, our result directly applies to an existing algorithm for the \cpugpu problem~\cite{DBLP:conf/wea/GermannJOT24}.
In this problem, we are again given a set of \(n\) jobs, but a job will be executed either on the GPU or the CPU.
On the GPU a job can be executed on a single machine and with a given processing time (this corresponds to $P||C_{max}$) while on the CPU a job is given a processing time function as in \ac{cpu}.
The task is now to find a partition of the jobs into two sets as well as a schedule for both sets such that the overall makespan is minimized.

While the GPU scheduling problem (\(P\|C_{\max}\)) has been researched continuously since the 1960s~\cite{DBLP:journals/siamam/Graham69,DBLP:journals/jacm/HochbaumS87,DBLP:journals/siamdm/Jansen10,DBLP:journals/siamcomp/CoffmanGJ78}, the combination of this problem with the more complex \ac{cpu} problem was first studied by Bleuse et al.~\cite{DBLP:journals/concurrency/BleuseKMMT15}.
They presented a (\(\frac{3}{2} + \varepsilon\))-approximation algorithm with running time \(O(n^2m_c^4m_g^2)\).
Later, Germann~et~al.~\cite{DBLP:conf/wea/GermannJOT24} improved the running time to \(O(\log(nm_cm_g/\varepsilon) \cdot n^2m_cm_g)\) while maintaining the same approximation ratio.
We show that our result for \ac{cpu} can be seamlessly integrated into their algorithm, leading to an improved approximation ratio for the combined \cpugpu problem.

By adjusting the relaxation parameter in~\cite{DBLP:conf/wea/GermannJOT24} to precisely match our tighter CPU bound, our analysis seamlessly plugs into their \ac{mmckp} formulation.
This straightforward substitution improves their dual approximation framework, yielding an improved overall approximation ratio of $\frac{2\sqrt{e}}{2\sqrt{e}-1}+\varepsilon$ for the combined \cpugpu problem.

\begin{restatable}{theorem}{cpugputheorem}
    Let \(\varepsilon > 0\). There exists a \((\frac{2\sqrt{e}}{2\sqrt{e}-1}+\varepsilon)\)-approximation algorithm for \cpugpu that runs in time \(O(\log(nm_cm_g/\varepsilon) \cdot n^2m_cm_g)\).
\end{restatable}

\paragraph*{Organization of the Paper}
In \Cref{sec:cpu-analysis}, we prove our main result, which is an improved analysis of the algorithm presented in~\cite{ohnesorge}.
Complementing our analysis, we provide a lower bound example in~\Cref{app:instance}, demonstrating that the algorithm cannot achieve an approximation ratio better than $\frac{2\sqrt{e}}{2\sqrt{e}-1}$.
Finally, in~\Cref{sec:cpu_gpu}, we show how this result directly implies an improved algorithm for the \cpugpu extension.

\subsection{Preliminaries}
Since we provide an improved analysis for the algorithm in~\cite{ohnesorge}, we briefly summarize their approach and establish the necessary notation.
Utilizing the dual approximation framework pioneered by Hochbaum and Shmoys~\cite{DBLP:journals/jacm/HochbaumS87}, we may assume that we are given a makespan guess \(d\) such that \(\opt \leq d < (1+\varepsilon)\opt \).
For simplicity, we also assume that $d=1$ (by scaling).

Given this, the algorithm in \cite{ohnesorge} partitions the jobs into a set of big jobs \(J_B \coloneqq \{j \in [n] \mid t_j(1) > \frac{3}{7}\}\) and a set of small jobs \(J_S = [n] \setminus J_B\) and focuses on scheduling the big jobs first.
The key component of the algorithm is partitioning the big jobs \(J_B\) into three sets using a \ac{mckp} formulation.
We restate the definition of the partitioning together with the result of our improved analysis in \Cref{lem:main-partition}, where \(\gamma_j(x)\) denotes the minimal machine assignment of job \(j\) such that its processing time is at most \(x\), i.e., \(\gamma_j(x) = \min\{\alpha \in [m] \mid t_j(\alpha) \leq x\}\).

\begin{lemma}[Improvement of~\cite{ohnesorge}]
    \label{lem:main-partition}
    Given a partition of jobs \(C_1 \sqcup C_2 \sqcup C_3 = J_B\) s.t.
    \begin{enumerate}
        \item \(\sum_{j \in C_1} w_j(\gamma_j(1)) + \sum_{j \in C_2} w_j(\gamma_j(\frac{4}{7})) + \sum_{j \in C_3} w_j(\gamma_j(\frac{3}{7})) \leq m - \sum_{j \in J_S} t_j(1)\)
        \item \(\sum_{j \in C_1} \gamma_j(1) + \frac12\sum_{j \in C_2} \gamma_j(\frac{4}{7}) \leq m\),
    \end{enumerate}
    one can compute a contiguous schedule with makespan at most \(\frac{2\sqrt{e}}{2\sqrt{e}-1}\) in time \(O(nm)\).
\end{lemma}
Note that the partition required for \Cref{lem:main-partition} can be found by a \ac{mckp}.
\Cref{alg:jansen_ohnesorge} gives a high-level overview on how the jobs are scheduled in \cite{ohnesorge}.
\begin{algorithm}[htpb]
    \caption{CMMJS Approximation Algorithm \cite{ohnesorge}}
    \label{alg:jansen_ohnesorge}
    \begin{algorithmic}[1]
        \Require Partition \(C_1 \sqcup C_2 \sqcup C_3 = J_B\) and small jobs \(J_S\)
        \State Construct \(S_0,S_1, S_2\) from \(C_1, C_2, C_3\) as in \cite{ohnesorge}. \Comment{\Cref{alg:assign_partitions}}

        \State Schedule jobs in $S_0\sqcup S_1$ at time $0$ so that the load of the machines is non-decreasing.
        \State Schedule jobs in $S_2$ with \cref{alg:RepairS2} or \cref{alg:assignment}.
        \State Greedily schedule small jobs \(J_S\).
    \end{algorithmic}
\end{algorithm}
For completeness, we include the procedure for constructing \(S_0, S_1\), and \(S_2\) in~\Cref{sec:assign_partitions} (\Cref{alg:assign_partitions}).
Here, we give only the high-level idea and summarize important properties of these sets in~\Cref{thm:base_properties}.



\begin{figure}[htbp]
    \centering
    \begin{subfigure}[b]{0.48\textwidth}
        \centering
        \begin{tikzpicture}[xscale=0.4, yscale=0.9, font=\sffamily]

            \def\m{8} 
            \def\opt{2} 
            \def\mylambda{1.41}
            \def\mylambdaprime{1.25}
            \def\mmax{12} 

            \pgfmathsetmacro{\lamopt}{\mylambda * \opt}
            \pgfmathsetmacro{\wzero}{\mylambdaprime * \m} 

            \draw[thick] (0, 0) -- (\mmax, 0) node[right] {};
            \draw[thick] (0, 0) -- (0, \lamopt) node[above=0.15cm] {};

            \draw[thick] (\m, 0.1) -- (\m, -0.1) node[below] {$m_1$};
            \draw[thick] (\mmax, 0.1) -- (\mmax, -0.1) node[below] {$m$};

            \draw[thick] (0.1, \opt) -- (-0.1, \opt) node[left] {$1$};
            \draw[thick] (0.1, \lamopt) -- (-0.1, \lamopt) node[left] {$\lambda$};
            \draw[thick] (0.1, {\lamopt/2}) -- (-0.1, {\lamopt/2}) node[left, font=\scriptsize] {$\lambda /2$};

            \draw[dashed, gray!60] (0, {\lamopt/2}) -- (\mmax, {\lamopt/2});
            \draw[dashed, gray!60] (0, \opt) -- (\mmax, \opt);
            \draw[dashed, gray!60] (0, \lamopt) -- (\mmax, \lamopt);

            \draw[dashed, gray!60] (\m, 0) -- (\m, \lamopt);
            \draw[dashed, gray!60] (\mmax, 0) -- (\mmax, \lamopt);

            \filldraw[fill=orange!20, draw=orange!80!black, thick] (7, 0) rectangle (8, 1.85);
            \filldraw[fill=orange!20, draw=orange!80!black, thick] (6, 0) rectangle (7, 1.75);
            \filldraw[fill=orange!20, draw=orange!80!black, thick] (5, 0) rectangle (6, 1.65);
            \filldraw[fill=orange!20, draw=orange!80!black, thick] (4, 0) rectangle (5, 1.55);
            \filldraw[fill=orange!20, draw=orange!80!black, thick] (3, 0) rectangle (4, 1.48);
            \filldraw[fill=orange!20, draw=orange!80!black, thick] (2.5, 0) rectangle (3, 1.34);

            \draw[|-|, thick] (0, -0.3) -- (2.5, -0.3) node[midway, below] {$q$};

            \filldraw[fill=teal!20, draw=teal!80!black, thick] (8, 0) rectangle (9, 2.1);
            \filldraw[fill=teal!20, draw=teal!80!black, thick] (9, 0) rectangle (10, 2.2);
            \filldraw[fill=teal!20, draw=teal!80!black, thick] (10, 0) rectangle (11, 2.3);
            \filldraw[fill=teal!20, draw=teal!80!black, thick] (11, 0) rectangle (12, 2.4);

        \end{tikzpicture}
        \caption{Green jobs are in $S_0$, orange jobs are in $S_1$.}
        \label{subfig:modified_problem}
    \end{subfigure}%
    \hfill%
    \begin{subfigure}[b]{0.48\textwidth}
        \centering
        \begin{tikzpicture}[xscale=0.4, yscale=0.9, font=\sffamily]

            \def\m{8} 
            \def\opt{2} 
            \def\mylambda{1.41}
            \def\mylambdaprime{1.25}
            \def\mmax{12} 

            \pgfmathsetmacro{\lamopt}{\mylambda * \opt}
            \pgfmathsetmacro{\wzero}{\mylambdaprime * \m} 

            \draw[thick] (0, 0) -- (\mmax, 0) node[right] {};
            \draw[thick] (0, 0) -- (0, \lamopt) node[above=0.15cm] {};

            \draw[thick] (\m, 0.1) -- (\m, -0.1) node[below] {$m_1$};
            \draw[thick] (\mmax, 0.1) -- (\mmax, -0.1) node[below] {$m$};

            \draw[thick] (0.1, \opt) -- (-0.1, \opt) node[left] {$1$};
            \draw[thick] (0.1, \lamopt) -- (-0.1, \lamopt) node[left] {$\lambda$};
            \draw[thick] (0.1, {\lamopt/2}) -- (-0.1, {\lamopt/2}) node[left, font=\scriptsize] {$\lambda /2$};

            \draw[dashed, gray!60] (0, {\lamopt/2}) -- (\mmax, {\lamopt/2});
            \draw[dashed, gray!60] (0, \opt) -- (\mmax, \opt);
            \draw[dashed, gray!60] (0, \lamopt) -- (\mmax, \lamopt);

            \draw[dashed, gray!60] (\m, 0) -- (\m, \lamopt);
            \draw[dashed, gray!60] (\wzero, 0) -- (\wzero, \lamopt);
            \draw[dashed, gray!60] (\mmax, 0) -- (\mmax, \lamopt);

            \filldraw[fill=teal!20, draw=teal!80!black, thick] (8, 0) rectangle (9, 2.1);
            \filldraw[fill=teal!20, draw=teal!80!black, thick] (9, 0) rectangle (10, 2.2);
            \filldraw[fill=teal!20, draw=teal!80!black, thick] (10, 0) rectangle (11, 2.3);
            \filldraw[fill=teal!20, draw=teal!80!black, thick] (11, 0) rectangle (12, 2.4);

            \filldraw[fill=blue!10, draw=blue!80!black, thick, opacity=0.7] (0, \opt) rectangle (\wzero, \lamopt);

            \filldraw[fill=orange!20, draw=orange!80!black, thick] (7, 0) rectangle (8, 1.85);
            \filldraw[fill=orange!20, draw=orange!80!black, thick] (6, 0) rectangle (7, 1.75);
            \filldraw[fill=orange!20, draw=orange!80!black, thick] (5, 0) rectangle (6, 1.65);
            \filldraw[fill=orange!20, draw=orange!80!black, thick] (4, 0) rectangle (5, 1.55);
            \filldraw[fill=orange!20, draw=orange!80!black, thick] (3, 0) rectangle (4, 1.48);
            \filldraw[fill=orange!20, draw=orange!80!black, thick] (2.5, 0) rectangle (3, 1.34);

            \draw[|-|, thick] (0, -0.3) -- (2.5, -0.3) node[midway, below] {$q$};

        \end{tikzpicture}
        \caption{The blue job is in $S_2$}
        \label{subfig:original_problem}
    \end{subfigure}

    \caption{(a) The schedule as in \cref{rem:schedule}. (b) The blue job cannot be placed on top of the orange jobs since its machine requirement is greater than $m_1$ and therefore intersects $S_0$.}
    \label{fig:job_scheduling_comparison}
\end{figure}
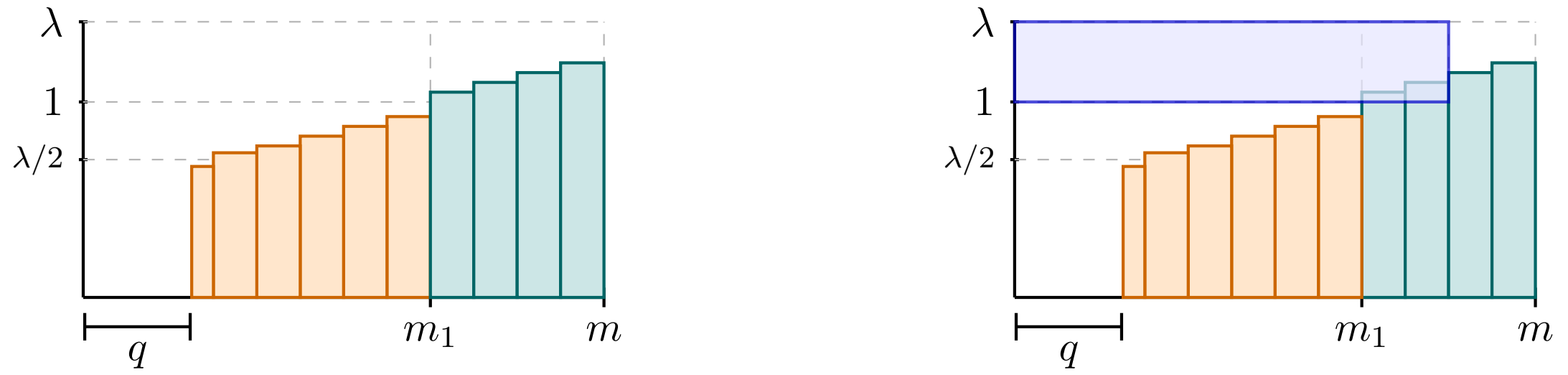

\begin{remark}[schedule]\label{rem:schedule}
    For a given target makespan $\lambda\in(1,\frac32)$, the original algorithm assigns each job $j\in S_0$ to $\alpha_j\coloneqq \gamma_j(\lambda)$ machines, each job \(j \in S_1\) to \(\alpha_j \coloneqq \gamma_j(1)\)  machines and each job \(j \in S_2\) to \(\alpha_j \coloneqq \gamma_j((\lambda - 1))\) machines, as illustrated in \Cref{subfig:original_problem}.
    In the following, we consider the jobs in $S_0$ and $S_1$ as already assigned to machines such that they are non-decreasing in processing time.
    Note that thereby some machines might not have a single job scheduled.
    We denote the number of machines without a scheduled job with $q\geq0$.
    This is depicted in \Cref{subfig:modified_problem}.
    The objective is to find a minimal $\lambda$, so that we can always compress the jobs in $S_2$ so that the schedule has a makespan of at most $\lambda$.
\end{remark}

Let $m_1 \coloneqq m - \sum_{j \in S_0} \gamma_j(\lambda)$ denote the number of machines remaining after placing $S_0$.
By construction, every job in $S_0$ has a processing time strictly greater than $1$.
Thus, the combined work area of the jobs in $S_1$ and $S_2$ is upper-bounded by $m_1$.
The result of these transformations is a so-called $3$-Shelf Schedule with the following properties:
\begin{lemma}[\cite{ohnesorge}, Definition 9 and Lemma 10]
    \label{thm:base_properties}
    Given $\lambda \in [\frac{10}{7}, \frac{3}{2})$ and a partition of \(C_1 \sqcup C_2 \sqcup C_3 = J_B\) as in \Cref{lem:main-partition}, we can return a contiguous schedule with a makespan of at most $\lambda$ in time $O(mn)$ or construct a partition \(S_0\sqcup S_1 \sqcup S_2 = J_B\) s.t.
    \begin{enumerate}
        \item
              \begin{enumerate}
                  \item The total work area is at most \(\sum_{j \in S_1} w_j(\gamma_j(1)) + \sum_{j \in S_2} w_j(\gamma_j(\lambda - 1)) \leq m_1\) \label[property]{cond:total_area}
              \end{enumerate}
        \item 
              \begin{enumerate}
                  \item All jobs in \(S_1\) can be scheduled on \(m_1 - q\) machines with makespan at most \(1\).
                  \item With the exception of one job assigned to a single machine, every job in $S_1$ has a processing time greater than $\frac\lambda2$.
                        \label[property]{cond:shelf1_height}
                  \item The total work area of $S_1$ is at least \(W_1 \coloneqq \sum_{j \in S_1} w_j(\gamma_j(1))\geq \frac\lambda2(m_1-q)\).
                        \label[property]{cond:shelf1_area}
              \end{enumerate}
        \item 
              \begin{enumerate}
                  \item Each job \(j \in S_2\) has work area at least \(w_j(\gamma_j((\lambda - 1))) > \lambda q \geq \lambda \) (since $q\geq 1$).\label[property]{cond:shelf2_area}
              \end{enumerate}
    \end{enumerate}
\end{lemma}

\section{A Better Analysis for \acl{cpu}}
\label{sec:cpu-analysis}

In this section we will assume that the properties given by \cref{thm:base_properties} hold since otherwise we can already find a schedule with makespan at most $\lambda$.
We will now place the jobs in $S_0\sqcup S_1$ as in \cref{alg:jansen_ohnesorge} and show how to handle $S_2$.


\begin{restatable}{lemma}{StwoEQone}
    \label{lem:s2_qeq_1}
    Let $\lambda \geq \frac{2\sqrt{e}}{2\sqrt{e}-1}$. If $|S_2|>1$, then we can create a contiguous schedule with makespan at most $\lambda$.
\end{restatable}
\begin{proof}
    We defer the proof of this lemma to \Cref{app:only_one_job}, since it is analogous to~\cite{ohnesorge}.
\end{proof}

Using the above lemma, we may assume $|S_2|=1$ for the remainder of the analysis.
We focus on the job in \(S_2\) and refer to it as $\job$ and to denote its work area as $\Aj$.
As in~\cite{ohnesorge}, we schedule $\job$ using the intuitive procedure outlined in \cref{alg:assignment}.
We compress $\job$ onto any possible number of machines in $[m_1]$ and place $\job$ with minimal makespan.

\begin{algorithm}[ht]
    \caption{Schedule Assignment}
    \begin{algorithmic}[1] 
        \label{alg:assignment}

        \Require Schedule from \cref{rem:schedule}. 
        \State $i_{min}=\arg\min_{i\in [m_1]}\load(i)+\frac{\Aj}{i}$
        \State Compress $\job$ on $i_{min}$ machines and execute it as early as possible on machines $1,\ldots,i_{min}$.
    \end{algorithmic}
\end{algorithm}

In the following, we provide a new and tight analysis for the outlined algorithm (specifically~\Cref{alg:assignment}).
Observe that \Cref{lem:s2_qeq_1,lem:s2_eq_1} together imply \Cref{thm:full}.
Therefore, the remainder of this section is dedicated to proving \Cref{lem:s2_eq_1}.
We will prove the lemma by carefully analyzing of the total work area of all jobs in $S_1$ and of the job $\job$ in $S_2$.

\begin{lemma}
    \label{lem:s2_eq_1}
    Let \(\lambda \geq \frac{2\sqrt{e}}{2\sqrt{e}-1}\).
    If $|S_2|=1$, then \Cref{alg:assignment} finds a contiguous monotone moldable job schedule with makespan at most \(\lambda\).
\end{lemma}
\begin{proof}
    We denote the single job in \(S_2\) as \(\job\).
    Note that \(\gamma_{\job}(\lambda - 1) > m_1\). Otherwise the job \(\job\) can be executed on machines \(1, \ldots, m_1\), resulting in a makespan of at most \(\lambda\) (see \Cref{fig:job_scheduling_comparison}).
    Therefore, we assume in the following that \(\gamma_{\job}(\lambda - 1) > m_1\).
    Since we denote by \(\Aj \coloneqq \sum_{j \in S_2} w_j(\gamma_j(\lambda - 1))\) the total work in \(S_2\), this implies:
    \begin{equation}\label{eq:work_job_lowerbound}
        \Aj>(\lambda-1)m_1.
    \end{equation}
    Combining \Cref{cond:total_area,cond:shelf1_area} gives us:
    \begin{equation}\label{eq:work_job_upperbound}
        \Aj\leq m_1-\frac{\lambda}{2}(m_1-q)=(1-\frac{\lambda}{2})m_1+\frac{\lambda}{2}q.
    \end{equation}
    Lastly, by interpolating between \Cref{eq:work_job_lowerbound,eq:work_job_upperbound}, we obtain that there exists a $c\in(0,1]$ that satisfies:
    \begin{equation}
        \label{eq:work-job-interpolation}
        \Aj = (1-c)(\lambda-1)m_1+c(1-\frac\lambda2)m_1+c\frac\lambda2q
    \end{equation}

    \begin{claim}\label{lem:q_upper_bound}
        The number of unused machines \(q\) is bounded by
        \begin{align}
            \label{eq:def_Q}
            q\leq Q(c) \coloneqq \frac{(1-c)(\lambda -1)m_1+c(1-\frac\lambda2)m_1}{(1-\frac{c}2)\lambda}.
        \end{align}

    \end{claim}
    \begin{claimproof}
        Assume the contrary, then we obtain:
        \begin{align*}
                                 & q> \frac{(1-c)(\lambda -1)m_1+c(1-\frac\lambda2)m_1}{(1-\frac{c}2)\lambda}                                                     \\
            \Leftrightarrow\quad & (1-\frac{c}{2})\lambda q> (1-c)(\lambda -1)m_1+c(1-\frac\lambda2)m_1                                                           \\
            \Leftrightarrow\quad & \lambda q>(1-c)(\lambda -1)m_1+c(1-\frac\lambda2)m_1+c\frac\lambda2 q                                                          \\
            \Leftrightarrow\quad & \lambda q>\Aj.                                                             &  & \text{using (\ref{eq:work-job-interpolation})}
        \end{align*}
        This is a direct contradiction to \Cref{cond:shelf2_area}.
    \end{claimproof}
    Now, we define a critical number of machines \(m'\coloneq \frac{2\Aj}{\lambda}\) so that for any \(\hat m\geq m'\):
    \begin{equation}\label{eq:m'_condition}
        \frac\lambda2-\frac{\Aj}{\hat m}\geq 0.
    \end{equation}
    Intuitively, this critical threshold is defined such that \(\job\) has a processing time of at most \(\frac{\lambda}{2}\) when it is assigned to more than \(m'\) machines.
    Next, we note that the single job with a processing time of at most \(\frac{\lambda}{2}\) from \Cref{cond:shelf1_height} results in a few edge cases that we need to consider separately.
    \Cref{prop:lem_sufficient_condition} provides the necessary conditions to handle these edge cases, which we will use in the following analysis.
    \begin{restatable}[Edge Cases]{claim}{edgecases}\label{prop:lem_sufficient_condition}
        \renewcommand{\theenumi}{\theclaim.\arabic{enumi}}
        \renewcommand{\labelenumi}{\arabic{enumi}.}
        \crefalias{enumi}{claim}
        Let $\lambda \in [\frac{10}7, \frac32]$ then:
        \begin{enumerate}
            \item \label{claim:load_assumption} If $q+1< \lceil m'\rceil -1$, then w.l.o.g. $\sum_{i=q+1}^{\lceil m'\rceil-1}\load(i)\geq \frac\lambda2(\lceil m'\rceil -q -1)$ holds.

            \item \label{claim:assignment_ratio} If $q+1=\lceil m'\rceil -1$, then \cref{alg:assignment} returns a schedule with makespan at most $\lambda$.

        \end{enumerate}
    \end{restatable}
    \begin{claimproof}
        To improve readability, we defer the proof to \Cref{sec:edge_case}.
    \end{claimproof}

    For the analysis, we define a function \(f(q)\), depending on the instance and \(q\):
    \begin{equation}
        \label{eq:def_f}
        f(q)\coloneqq (\lambda-1)m_1-\frac\lambda2q-\Aj\ln\left(\frac{m_1}{m'}\right).
    \end{equation}
    Note that for any specific, fixed instance, \(\Aj\) and \(m'\) are constants, allowing us to analyze \(f\) strictly as a function of the variable \(q\).
    First, we will show that \(f(q) \geq 0\) implies that \cref{alg:assignment} returns a schedule with makespan at most \(\lambda\).
    Later, in \Cref{lem:f_geq_0}, we will show that \(f(q) \geq 0\) holds for all instances with \(\lambda \in [\frac{10}{7}, \frac{3}{2}]\).
    To prepare the proof we list the following analytical properties:
    \begin{restatable}[Analytical Properties]{claim}{anaproperties}
        \renewcommand{\theenumi}{\theclaim.\arabic{enumi}}
        \renewcommand{\labelenumi}{\arabic{enumi}.}
        \crefalias{enumi}{claim}
        Let $\lambda \in [\frac{2\sqrt{e}}{2\sqrt{e}-1},\frac32]$, then:
        \begin{enumerate}
            \item \label{claim:sum_greater_int} $\sum_{i=\lceil{m'}\rceil}^{m_1}\left[\frac\lambda2-\frac{\Aj}i\right]>\int_{m'}^{m_1}\frac\lambda2-\frac{\Aj}x\ dx $.
            \item \label{claim:f_decreasing} $f(q)=(\lambda-1)m_1-\frac\lambda2q-\Aj\ln(\frac{m_1}{m'})$ monotonically decreases in $q$.
            \item \label{claim:q_increasing} $\qc= \frac{(1-c)(\lambda -1)m_1+c(1-\frac\lambda2)m_1}{(1-\frac{c}2)\lambda}$ monotonically increases in $c$.
            \item \label{claim:g} $g(\qc)=(\lambda-1)m_1-\frac{\lambda}{2}\qc-\lambda \qc\ln\frac{m_1\lambda}{2\lambda \qc}$ monotonically increases in $\qc$.
        \end{enumerate}
    \end{restatable}
    \begin{claimproof}
        To improve readability, we defer the proof to \Cref{sec:ana_proofs}.
    \end{claimproof}

    \begin{claim}\label{lem:sufficient_condition}
        Let $\lambda \in [\frac{10}7,\frac32]$. If $f(q)\geq 0$, then \cref{alg:assignment} returns a contiguous schedule with a makespan at most $\lambda$.
    \end{claim}
    \begin{claimproof}
        By \Cref{cond:total_area}, we have for the work area of \(S_1\): $W_1 \leq m_1-\Aj$. By definition of $f$ (see \cref{eq:def_f}), we have:
        \begin{align}
                                 & (\lambda-1)m_1-\frac\lambda2q-\Aj\ln(\frac{m_1}{m'})\geq 0                                          \nonumber \\
            \Leftrightarrow\quad & (\lambda-1)m_1-\frac{\lambda}{2}\frac{2}{\lambda}\Aj-\frac\lambda2q-\Aj\ln(\frac{m_1}{m'})\geq -\Aj \nonumber \\
            \Leftrightarrow\quad & \frac\lambda2(m_1-q)+\frac{\lambda}{2}(m_1-\frac{2}{\lambda}\Aj)-\Aj\ln(\frac{m_1}{m'})\geq m_1-\Aj \nonumber \\
            \Rightarrow\quad     & \frac\lambda2(m_1-q)+\frac{\lambda}{2}(m_1-m')-\Aj\int_{m'}^{m_1}\frac{1}{x}\ dx\geq W_1            \nonumber \\
            \Leftrightarrow\quad & \frac\lambda2(m_1-q)+\int_{m'}^{m_1}\frac{\lambda}{2}-\frac{\Aj}{x}\ dx\geq W_1. \label{eq:amax}
        \end{align}
        With \Cref{claim:sum_greater_int}, we get:
        \begin{align}\label{eq:amax_sum}
             & \frac\lambda2(m_1-q)+\sum_{i=\lceil m'\rceil}^{m_1}\left[\frac{\lambda}{2}-\frac{\Aj}{i}\right]> W_1.
        \end{align}
        Furthermore, since the jobs in $S_1$ are scheduled non-decreasingly (see \cref{rem:schedule}):
        \begin{align}\label{eq:amax_upperbound}
            W_1
            = \sum_{i=1}^{m_1}\load(i)
            = \sum_{i=q+1}^{\lceil m'\rceil-1}\load(i)+\sum_{i=\lceil m'\rceil}^{m_1}\load(i).
        \end{align}

        \noindent\textbf{Case 1 ($q+1=\lceil m'\rceil -1$):}
        In this case we can directly apply \Cref{claim:assignment_ratio} to conclude that \cref{alg:assignment} returns a schedule with makespan at most $\lambda$.

        \noindent\textbf{Case 2 (\(q+1<\lceil m'\rceil -1\)):}
        In this case we can use \Cref{claim:load_assumption} and \cref{eq:amax_upperbound} to bound
        \begin{equation}
            \label{eq:amax_q_small}
            W_1 \geq \frac\lambda2\left(\lceil m' \rceil-q-1\right)+\sum_{i=\lceil m'\rceil}^{m_1}\load(i).
        \end{equation}

        Combining \cref{eq:amax_sum,eq:amax_q_small} yields:
        \begin{align*}
                                 & \frac\lambda2(m_1-q)+\sum_{i=\lceil m'\rceil}^{m_1}\left[\frac{\lambda}{2}-\frac{\Aj}{i}\right]
                                 & >                                                                                                               & \frac\lambda2\left(\lceil m' \rceil-q-1\right)+\sum_{i=\lceil  m'\rceil}^{m_1}\load(i) \\
            \Leftrightarrow\quad & \frac\lambda2(m_1-\lceil m'\rceil+1)+\sum_{i=\lceil m'\rceil}^{m_1}\left[\frac{\lambda}{2}-\frac{\Aj}{i}\right]
                                 & >                                                                                                               & \sum_{i=\lceil m'\rceil}^{m_1}\load(i)                                                 \\
            \Leftrightarrow\quad & \sum_{i=\lceil m'\rceil}^{m_1}\left[\lambda-\frac{\Aj}{i}\right]
                                 & >                                                                                                               & \sum_{i=\lceil m'\rceil}^{m_1}\load(i)
        \end{align*}
        This implies that there exists an $i'\in [\lceil m'\rceil,m_1]$ such that:
        \begin{align*}
            \lambda-\frac{\Aj}{i'}> \load(i')
            \Leftrightarrow \lambda >\load(i')+\frac{\Aj}{i'}.
        \end{align*}
        Note that \(\frac{\Aj}{i'}\) corresponds to the maximum processing time of the single job \(\job \in S_2\), when compressing it from \(\gamma_{\job}(\lambda - 1)\) machines to \(i'\) machines.
        In other words, we can process the single job $\job \in S_2$ on $i'$ machines without increasing the makespan beyond $\lambda$, due to monotony assumptions.
        Note that machines $\{1,\ldots,i'-1\}$ have a load of at most $\load(i')$, and therefore the job fits there as well.

        \noindent\textbf{Case 3 ($q+1>\lceil m'\rceil -1$):}
        In this case $\sum_{i=q+1}^{\lceil m'\rceil-1}\load(i)=0$ holds. Thus, \cref{eq:amax_upperbound} simplifies to:
        \begin{equation}
            \label{eq:amax_q_large}
            W_1 = \sum_{i=\lceil m'\rceil}^{m_1}\load(i).
        \end{equation}
        Note that $q$ and $\lceil m'\rceil$ are integers.
        Therefore, $q\geq \lceil m'\rceil-1$ holds or equivalently $-\lceil m'\rceil +1\geq -q$.
        Combining this and \cref{eq:amax_sum,eq:amax_q_large} yields:
        \begin{align*}
                                 & \frac\lambda2(m_1-q)+\sum_{i=\lceil m'\rceil}^{m_1}\left[\frac{\lambda}{2}-\frac{\Aj}{i}\right]
                                 & >                                                                                                               & \sum_{i=\lceil  m'\rceil}^{m_1}\load(i) \\
            \implies\quad        & \frac\lambda2(m_1-\lceil m'\rceil+1)+\sum_{i=\lceil m'\rceil}^{m_1}\left[\frac{\lambda}{2}-\frac{\Aj}{i}\right]
                                 & >                                                                                                               & \sum_{i=\lceil m'\rceil}^{m_1}\load(i)  \\
            \Leftrightarrow\quad & \sum_{i=\lceil m'\rceil}^{m_1}\left[\lambda-\frac{\Aj}{i}\right]
                                 & >                                                                                                               & \sum_{i=\lceil m'\rceil}^{m_1}\load(i)
        \end{align*}
        Analogously to Case 2, this implies that there exists an $i'\in [\lceil m'\rceil,m_1]$ such that:
        \begin{align*}
            \lambda-\frac{\Aj}{i'}> \load(i')
            \Leftrightarrow \lambda >\load(i')+\frac{\Aj}{i'}.
        \end{align*}
        In other words, we can process the job $\job$ on $i'$ machines without increasing the makespan beyond $\lambda$.
        This completes the proof of \Cref{lem:sufficient_condition}.
    \end{claimproof}

    We have now shown that if \(f(q)\geq 0\), then \cref{alg:assignment} returns a schedule with makespan at most \(\lambda\).
    It remains to show that \(f(q)\geq 0\) for all valid values of \(q\).

    \begin{claim}\label{lem:f_geq_0}
        Let $\lambda \in [\frac{2\sqrt{e}}{2\sqrt{e}-1},\frac{3}{2}]$. Then, $f(q)= (\lambda-1)m_1-\frac\lambda2q-\Aj\ln(\frac{m_1}{m'})\geq 0$.
    \end{claim}
    \begin{claimproof}
        Note, that by \Cref{claim:f_decreasing}, we have that \(f(q)\) is monotonically decreasing in \(q\).
        Further, by \Cref{lem:q_upper_bound} we know that \(q \leq \qc = \frac{(1-c)(\lambda -1)m_1+c(1-\frac\lambda2)m_1}{(1-\frac{c}2)\lambda}\).
        Therefore, to prove \(f(q) \geq 0\), it is sufficient to show that $f(\qc)\geq 0$, and for the remainder of this proof, we may assume that \(q = \qc\).

        Let $Z \coloneqq (1-c)(\lambda-1)m_1+c(1-\frac{\lambda}{2})m_1$.
        By \Cref{eq:work-job-interpolation}, the work area equals:
        \begin{align*}
            \Aj = Z + c \frac{\lambda}{2}\qc.
        \end{align*}
        By definition: $Q(c) = \frac{Z}{(1 - \frac{c}{2})\lambda}\Leftrightarrow Z = \lambda Q(c) - c\frac{\lambda}{2}Q(c)$. By substitution:
        \begin{align}
            \label{eq:aj_equality}
            \Aj = \left( \lambda Q(c) - c\frac{\lambda}{2}Q(c) \right) + c \frac{\lambda}{2}Q(c) = \lambda Q(c).
        \end{align}
        Using \Cref{eq:aj_equality} and substituting $m' = \frac{2W_2}{\lambda}$, we can evaluate $f(Q(c))$ purely in terms of $Q(c)$:
        \begin{align}
            f(\qc)
             & = (\lambda-1)m_1-\frac{\lambda}{2}\qc-\Aj \ln \frac{m_1}{m'}                                                                 \nonumber                                          \\
             & = (\lambda-1)m_1-\frac{\lambda}{2}\qc-\Aj\ln \frac{m_1\lambda}{2\Aj}                                                        \nonumber                                           \\
             & =(\lambda-1)m_1-\frac{\lambda}{2}\qc-\lambda \qc\ln\frac{m_1\lambda}{2\lambda \qc}.                                                    &  & \text{using (\ref{eq:aj_equality})}
        \end{align}
        By evaluating the function at this worst-case boundary, the previously constant instance parameters become functionally dependent on \(\qc\).
        Let $g(Q(c))$ denote this evaluated expression:
        \begin{equation}
            \label{eq:def_gq}
            g(Q(c)) \coloneqq (\lambda-1)m_1-\frac{\lambda}{2}Q(c)-\lambda Q(c)\ln\left(\frac{m_1}{2Q(c)}\right).
        \end{equation}
        By \Cref{claim:q_increasing}, \(\qc\) increases monotonically in \(c\).
        This implies $\qc\geq \qz=\frac{\lambda-1}{\lambda}m_1$.
        Further, by \Cref{claim:g}, \(g(\qc)\) increases monotonically in \(\qc\). Thus:
        \begin{align*}
            g(\qc) \geq g(Q(0)).
        \end{align*}
        It remains to show that $g(Q(0)) \geq 0$. By our initial assumption, we have $\frac{2\sqrt{\epsilon}}{2\sqrt{\epsilon}-1} \leq \lambda$.
        We can rearrange this to bound the constant $\sqrt{e}$:
        \begin{alignat*}{3}
                                 &  & \frac{2\sqrt{e}}{2\sqrt{e}-1}                & {}\leq \lambda                                                                                             \\
            \Leftrightarrow\quad &  & 2\sqrt{e}                                    & {}\leq \lambda(2\sqrt{e}-1)                                                                                \\
            \Leftrightarrow\quad &  & 2\sqrt{e}                                    & {}\leq 2\lambda\sqrt{e}-\lambda                                                                            \\
            \Leftrightarrow\quad &  & \lambda                                      & {}\leq 2\sqrt{e}(\lambda-1)                                                                                \\
            \Leftrightarrow\quad &  & \frac{\lambda}{2(\lambda-1)}                 & {}\leq \sqrt{e}                                                                                            \\
            \intertext{Since $\lambda \geq \frac{2\sqrt{e}}{2\sqrt{e}-1} > 1$, both sides of the inequality are strictly positive. Thus, we can safely apply the natural logarithm:}
            \implies\quad        &  & \ln\left(\frac{\lambda}{2(\lambda-1)}\right) & {}\leq \frac{1}{2}                                                                                         \\
            \Leftrightarrow\quad &  & 0                                            & {}\leq \frac{1}{2}-\ln\left(\frac{\lambda}{2(\lambda-1)}\right)                                            \\
            \intertext{Next, we scale the inequality. Because $m_1 \geq 1$ and $\lambda > 1$, the term $(\lambda-1)m_1$ is strictly positive. Multiplying both sides by this term preserves the inequality:}
            \implies\quad        &  & 0                                            & {}\leq (\lambda-1)m_1\left(\frac{1}{2}-\ln\left(\frac{\lambda}{2(\lambda-1)}\right)\right)                 \\
            \Leftrightarrow\quad &  & 0                                            & {}\leq \frac{1}{2}(\lambda-1)m_1-(\lambda-1)m_1\ln\left(\frac{\lambda}{2(\lambda-1)}\right)                \\
            \Leftrightarrow\quad &  & 0                                            & {}\leq (\lambda-1)m_1-\frac{1}{2}(\lambda-1)m_1-(\lambda-1)m_1\ln\left(\frac{\lambda}{2(\lambda-1)}\right) \\
            \intertext{Finally, we substitute $\qz = \frac{\lambda-1}{\lambda}m_1$. Noting that $\lambda\qz = (\lambda-1)m_1$, we can rewrite the coefficients and simplify the fraction inside the logarithm to reconstruct the function $g(\qz)$:}
            \Leftrightarrow\quad &  & 0                                            & {}\leq (\lambda-1)m_1-\frac{\lambda}{2}\qz-\lambda \qz \ln\left(\frac{m_1}{2\qz}\right)                    \\
            \Leftrightarrow\quad &  & 0                                            & {}\leq g(\qz). \tag*{using (\ref{eq:def_gq})}
        \end{alignat*}
        This completes the proof of \Cref{lem:f_geq_0}.
    \end{claimproof}
    The combination of \Cref{lem:sufficient_condition} and \Cref{lem:f_geq_0} shows that \cref{alg:assignment} returns a schedule with makespan at most \(\lambda\) for all valid values of \(c\) and \(q\) if \(|S_2| = 1\).
\end{proof}

\Cref{lem:s2_qeq_1,lem:s2_eq_1} prove our main result.
For a proof of the running time, we refer to~\cite{ohnesorge}, since we only provided a new analysis and did not alter the algorithm.
\FullTheoremCmd*
The preceding analysis shows that for some given area of $\job$ there is a function (seen in \cref{fig:instance}), so that each job in the schedule must have a higher processing time at the corresponding position.
We show that the analysis is tight in \cref{app:instance} by constructing an instance which is a lower sum to this function and approaches it arbitrary close.

\section{Extension to CPU/GPU Scheduling}
\label{sec:cpu_gpu}

The problem of \cpugpu generalizes the previously discussed CPU scheduling by introducing a second set of $m_g$ identical GPU machines. Each job can be scheduled either as a moldable job on the CPU or sequentially on a single GPU machine.

To extend our improved ratio of $\lambda = \frac{2\sqrt{e}}{2\sqrt{e}-1}$ to the \cpugpu problem, we adopt the dual approximation framework and the \ac{mmckp} partitioning formulation introduced by Germann et al.~\cite{DBLP:conf/wea/GermannJOT24}.
We denote a makespan guess as $d \in \mathbb{Q}_{>0}$. To bound the schedule on the GPU, Germann et al.~\cite{DBLP:conf/wea/GermannJOT24} introduced a size function $\varphi(j)$ that classifies GPU jobs based on their processing time $p(j)$:
\begin{equation}
    \varphi(j) \coloneq \begin{cases}
        \max\{k \in \{0, 1, 2\} : \frac{kd}{3} \le p(j)\} & \text{if } p(j) \le d, \\
        +\infty                                           & \text{otherwise.}
    \end{cases}
\end{equation}

By limiting the total size of jobs assigned to the GPU partition ($\mathcal{T}_g$), we can guarantee that at most two jobs with a processing time greater than $\frac{d}{3}$ are scheduled on any single GPU machine. We restate the following core lemma which defines the relaxation variable $\delta$ required to match the GPU makespan to our CPU makespan.

\begin{lemma}[\cite{DBLP:conf/wea/GermannJOT24}]
    \label{lem:gpu_lpt}
    Let $d \in \mathbb{Q}_{>0}$ be a makespan guess, $\delta \in \mathbb{Q}_{>0}$, and $\mathcal{T}_g \subseteq [n]$ be the jobs assigned to the GPU. If $\sum_{j \in \mathcal{T}_g} \varphi(j) \le 2m_g$ and the total work on the GPU satisfies $\sum_{j \in \mathcal{T}_g} p(j) \le m_g(1+\delta)d$, we can construct a GPU schedule of length at most $(\frac{4}{3} + \delta)d$ in time $O(n \log n)$ using \ac{lpt} scheduling.
\end{lemma}

To apply this to our new result, we need the GPU schedule to match our CPU ratio of $\lambda$. Therefore, we set our relaxation constant $\delta$ such that the GPU makespan equals our CPU makespan:
\begin{equation}
    \frac{4}{3} + \delta = \lambda \implies \delta = \frac{2\sqrt{e}}{2\sqrt{e}-1} - \frac{4}{3} \approx 0.1018
\end{equation}

With $\delta$ defined, we can find a valid partition of jobs across the CPU shelves ($S_0, S_1, S_2$) and the GPU ($\mathcal{T}_g$) by solving the \ac{mmckp}. To ensure the dynamic program runs in polynomial time, the GPU processing times are scaled and rounded as $\tilde{p}(j) \coloneq \lfloor \mu p(j) \rfloor$ with a scaling factor $\mu \coloneq \frac{n}{\delta d m_g}$.

\begin{lemma}[Knapsack Formulation]
    \label{lem:knapsack_cpu_gpu}
    Finding a partition that satisfies the CPU constraints established in \Cref{lem:main-partition} alongside the GPU constraints $\sum_{j \in \mathcal{T}_g} \varphi(j) \le 2m_g$ and $\sum_{j \in \mathcal{T}_g} \tilde{p}(j) \le \frac{n}{\delta}$ can be solved via dynamic programming in time $O(n^2 m_c m_g)$.
\end{lemma}
\begin{proof}
    The statement follows from a standard dynamic programming formulation for the \ac{mmckp} problem~\cite{DBLP:books/kellerer/knapsackproblems}. The details of the dynamic programming approach are analogous to those used in~\cite{DBLP:conf/wea/GermannJOT24}.
\end{proof}

\cpugputheorem*
\begin{proof}
    Using the framework from Lemma~\ref{lem:knapsack_cpu_gpu}, we can compute a valid job partition across the CPU and GPU in $O(n^2 m_c m_g)$ time.
    For the CPU partition, our algorithm from Section 4 guarantees a schedule of length at most $\lambda d = \frac{2\sqrt{e}}{2\sqrt{e}-1}d$. For the GPU partition, the knapsack constraints ensure that $\sum_{j \in \mathcal{T}_g} \varphi(j) \le 2m_g$ and the total relaxed workload is bounded by $m_g(1+\delta)d$. By Lemma~\ref{lem:gpu_lpt}, the \ac{lpt} algorithm will construct a valid GPU schedule of length at most $(\frac{4}{3} + \delta)d = \lambda d$ in time $O(n \log n)$.

    Using the primal approximation framework, we can use a binary search procedure with $O(\log(nm_cm_g/\varepsilon))$ iterations.
    Since the running time is strictly dominated by solving the \ac{mmckp} in each iteration, the total time complexity is bounded by $O(\log(nm_cm_g/\varepsilon) \cdot n^2m_cm_g)$.
\end{proof}

\section{Conclusion}

In this paper, we provide an improved analysis for the \acl{cpu} (\acs{cpu}) problem.
With a careful geometric argument, we successfully reduced the approximation ratio from approximately $1.459+\varepsilon$ \cite{ohnesorge} to $\frac{2\sqrt{e}}{2\sqrt{e}-1}+\varepsilon \approx 1.435+\varepsilon$. Furthermore, we established that our worst-case analysis is essentially tight for this algorithmic approach by providing a matching lower bound instance. We also demonstrated the versatility of this algorithm by seamlessly extending it to the heterogeneous \cpugpu problem, achieving an identical approximation guarantee.

Despite these improvements, several natural open questions remain. A primary theoretical focus for future research is determining the true lower bound for practically efficient algorithms in this setting. Since the structural properties of our partition (\cref{thm:base_properties}) hold for any $\lambda \in [\frac{10}{7}, \frac{3}{2}]$, it remains an intriguing open question whether \cref{alg:assignment} can be augmented---or a novel assignment strategy developed---to push the approximation ratio down to the $\frac{10}{7} \approx 1.428$ barrier.

Finally, another important direction is to further improve the running time. While our algorithm is practically efficient with a time complexity of $O(nm \log \frac{1}{\varepsilon})$, reducing the dependency on the number of machines $m$, similar to~\cite{DBLP:conf/europar/GrageJO23,land2013boundingmoldable}, would make the algorithm effectively linear in \(n\), while our running time is currently quadratic in \(n\) (if \(m \approx n\)).

\printbibliography

\appendix
\crefalias{section}{appendix}
\section{There is at most one job in $S_2$}
In this section, we will prove the following lemma.
\label{app:only_one_job}

\StwoEQone*

To prove this lemma, we make a case distinction on the value of \(q\).
If \(q\) is large, we can directly show that there can be at most one job in \(S_2\) by using the area constraints.
If \(q\) is small, we can show that a feasible schedule can be found by \Cref{alg:RepairS2}.
As stated before, this procedure is identical to the one used in \cite{ohnesorge}.

\begin{observation} \label{obs:single_job_large_q}
    Let $\lambda \geq \frac{2\sqrt{e}}{2\sqrt{e}-1}$. If $q>\frac{1}{3}(1-\frac{1}{\sqrt{e}})m_1$, then $|S_2|\leq 1$.
\end{observation}
\begin{proof}
    Assume the opposite, i.e., $|S_2|>1$. By \Cref{cond:total_area,cond:shelf1_area,cond:shelf2_area}, we have:
    \begin{align*}
        m_1 & \overset{\ref{cond:total_area}}\geq \sum_{j\in S_1}w_j(\gamma_j(1)) + \Aj                                                                                                  \\
            & > \frac{\lambda}{2}(m_1-q) + 2\lambda                                                                      &  & (|S_2|>1, \text{\Cref{cond:shelf1_area,cond:shelf2_area}}) \\
            & = \frac{\lambda}{2}m_1 + \frac{3}{2}\lambda q                                                                                                                              \\
            & = \frac{\lambda}{2}m_1 + \frac{3}{2}\lambda \left( \frac{1}{3}\left(1-\frac{1}{\sqrt{e}}\right)m_1 \right)                                                                 \\
            & = \frac{\lambda}{2}m_1 + \frac{\lambda}{2}\left(1-\frac{1}{\sqrt{e}}\right)m_1                                                                                             \\
            & = \frac{\lambda}{2}m_1 \left( 2 - \frac{1}{\sqrt{e}} \right)                                                                                                               \\
            & = \lambda m_1 \left( \frac{2\sqrt{e}-1}{2\sqrt{e}} \right).
    \end{align*}

    Comparing the first and last terms and dividing by $m_1$, we get:
    \begin{equation*}
        1 > \lambda \left( \frac{2\sqrt{e}-1}{2\sqrt{e}} \right) \implies \lambda < \frac{2\sqrt{e}}{2\sqrt{e}-1}.
    \end{equation*}
    This is a contradiction.
\end{proof}

We now show that if \(q\) is small, we can find a feasible schedule with \Cref{alg:RepairS2}.

\begin{algorithm}[H]
    \caption{RepairS2}\label{alg:RepairS2}
    \begin{algorithmic}[1]
        \While {$m_2 > m_1$}
        \State Let $j$ be the job with the smallest processing time among jobs in $S_2$.
        \State Let $p$ be the number of machines $j$ is allotted to.
        \State Allot $j$ to $p - 1$ machines.
        \EndWhile
        \State Sort jobs in $S_1$ in descending order of execution time.
        \State Sort jobs in $S_2$ in ascending order of execution time.
    \end{algorithmic}
\end{algorithm}

\begin{observation}
    \label{obs:single_job_small_q}
    Let $\lambda \geq \frac{2\sqrt{e}}{2\sqrt{e}-1}$. If $q\leq \frac{1}{3}(1-\frac{1}{\sqrt{e}})m_1$, then we can find a feasible schedule with \cref{alg:RepairS2}.
\end{observation}
\begin{proof}

    We prove this statement by analyzing \Cref{alg:RepairS2}.
    Denote in the following the total work area of jobs in \(S_1\) as \(W_1 \coloneqq \sum_{j \in S_1} w_j(\alpha_j)\) and the total work area of jobs in \(S_2\) as \(W_2 \coloneqq \sum_{j \in S_2} w_j(\alpha_j)\).

    \begin{claim}
        \label{obs:18}
        $W_1 > \frac{1+2\sqrt{e}}{6\sqrt{e}-3}m_1$ and $W_2<\frac{4\sqrt{e}-4}{6\sqrt{e}-3}m_1$.
    \end{claim}
    \begin{claimproof}
        By \Cref{cond:shelf1_area} we have:
        \begin{align*}
            W_1>\frac{\lambda}{2}(m_1-q)
            \geq \frac{1}{2}\frac{2\sqrt{e}}{2\sqrt{e}-1}(m_1-\frac{1}{3}(1-\frac{1}{\sqrt{e}})m_1)
            =\frac{1+2\sqrt{e}}{6\sqrt{e}-3}m_1
        \end{align*}
        And $W_1+W_2\leq m_1$ implies $W_2\leq m_1-W_1<m_1-\frac{1+2\sqrt{e}}{6\sqrt{e}-3}m_1=\frac{4\sqrt{e}-4}{6\sqrt{e}-3}m_1$.
    \end{claimproof}
    \begin{claim}
        After \Cref{alg:RepairS2} every job in \(S_2\) is executed on at least \(3\) machines.
        \label{obs:19}
    \end{claim}
    \begin{claimproof}
        Before the application of \Cref{alg:RepairS2}, every job $j \in S_2$ is executed on \(\gamma_j(\lambda - 1)\) machines. By the definition of $\gamma_j$, the processing time of each job is at most \(\lambda - 1 = \frac{1}{2\sqrt{e}-1}\).
        Furthermore, by \Cref{cond:shelf2_area}, every job in \(S_2\) has a work area of at least \(\lambda q \geq \frac{2\sqrt{e}}{2\sqrt{e}-1}\).
        These two properties imply that every job in $S_2$ is initially assigned to at least $\frac{\lambda q}{\lambda - 1} \ge 2\sqrt{e} > 3$ machines.

        Assume for the sake of contradiction that at some point during \Cref{alg:RepairS2}, a job $j$ is compressed from $3$ machines down to $2$ machines.
        Because \Cref{alg:RepairS2} always selects the job with the smallest processing time to compress, this implies that at this exact step, every remaining job in $S_2$ has a current processing time of at least $t_j(3)$.

        We can establish a strict lower bound for $t_j(3)$ by using the gap $q$ and the monotony properties of the jobs.
        Recall that no job in $S_2$ could be scheduled in the gap of $q$ machines with a processing time of at most $\lambda$ since the work must be at above $\lambda q$ and the machine allotment above $q$ to be placed in $S_2$.
        Thus, it must hold that $t_j(q) > \lambda$ by monotony
        We distinguish two cases based on the size of $q$:

        \noindent\textbf{Case 1 ($q \ge 3$):} We apply the monotony property that the processing time function $t_j$ is non-increasing. This directly yields $t_j(3) \ge t_j(q) > \lambda > \frac{\lambda}{3}$.

        \noindent\textbf{Case 2 ($q < 3$):} Since $q$ is an integer, $q \in \{1, 2\}$. We apply the monotony property that the work area function $w_j$ is non-decreasing. This gives $w_j(3) \ge w_j(q) = q \cdot t_j(q) > q\lambda \ge \lambda$. Consequently, the processing time on $3$ machines is $t_j(3) = \frac{w_j(3)}{3} > \frac{\lambda}{3}$.

        In both cases, we have that $t_j(3) > \frac{\lambda}{3}$. Therefore, at the moment of this compression, every job in $S_2$ has a processing time strictly greater than $\frac{\lambda}{3}$.

        Let $m_2$ be the total number of machines currently assigned to jobs in $S_2$. Since the \textproc{While}-loop in \Cref{alg:RepairS2} is still running, we know that $m_2 > m_1$.
        The current total work area of $S_2$ is strictly greater than $\frac{\lambda}{3} m_2 > \frac{\lambda}{3} m_1$.

        Because the work area of a job is non-decreasing with respect to the number of machines, compressing jobs to fewer machines never increases their total work area. Hence, the initial total work area $W_2$ must be at least the current total work area, meaning $W_2 > \frac{\lambda}{3} m_1$.
        Since $\lambda \ge \frac{2\sqrt{e}}{2\sqrt{e}-1}$, we have:
        \begin{align*}
            W_2 & > \frac{2\sqrt{e}}{3(2\sqrt{e}-1)} m_1 = \frac{2\sqrt{e}}{6\sqrt{e}-3} m_1.
        \end{align*}
        Because $e < 4$, it follows that $\sqrt{e} < 2$, which implies $2\sqrt{e} > 4\sqrt{e} - 4$. Thus, we obtain:
        \begin{align*}
            W_2 & > \frac{4\sqrt{e}-4}{6\sqrt{e}-3} m_1.
        \end{align*}
        This strictly contradicts \Cref{obs:18}, which bounded the area as $W_2 < \frac{4\sqrt{e}-4}{6\sqrt{e}-3} m_1$. We conclude that the algorithm never compresses a job to 2 machines.
    \end{claimproof}

    \begin{claim}
        \label{obs:20}
        Before \cref{alg:RepairS2} every job in $S_2$ has a processing time greater than $\frac{4}{10\sqrt{e}-5}$.
    \end{claim}
    \begin{claimproof}
        We make a case distinction on the number of machines \(\gamma_j(\lambda-1)\) that job \(j\) is executed on before \cref{alg:RepairS2}:

        \noindent\textbf{Case 1 (\(\gamma_j(\lambda-1) \leq 4\)):}
        Every job $j$ has an area of at least $\lambda$ by \Cref{cond:shelf2_area}.
        This implies that the processing time is at least $\frac\lambda4>\frac{4}{10\sqrt{e}-5}$.

        \noindent\textbf{Case 2 (\(\gamma_j(\lambda-1) > 4\)):}
        In this case we get:
        \begin{align*}
            h_j(\gamma_j(\lambda-1))  >(\lambda-1)\frac{\gamma_j(\lambda-1)-1}{\gamma_j(\lambda-1)}
            \geq (\lambda-1)\frac45 = \frac{4}{10\sqrt{e}-5}.
        \end{align*}
        The first inequality holds by the definition of $\gamma_j$, i.e., \(\gamma_j(x) = \min\{\alpha \in [m] \mid h_j(\alpha) \leq x\}\).
    \end{claimproof}

    \begin{claim}
        After \Cref{alg:RepairS2}, every job in $S_2$ has a processing time of at most $\frac{4}{3}\frac{4\sqrt{e}-4}{6\sqrt{e}-3}$.
    \end{claim}
    \begin{claimproof}
        Assume for the sake of contradiction that there exists at least one job in $S_2$ with a processing time strictly greater than $\frac{4}{3}\frac{4\sqrt{e}-4}{6\sqrt{e}-3}$.

        Recall that \Cref{alg:RepairS2} iteratively compresses the job with the smallest current processing time.
        Because the algorithm consistently selects the job with the minimum processing time to compress, if the current minimum processing time exceeds this threshold, then all jobs currently in $S_2$ must possess a processing time of at least $\frac{4}{3}\frac{4\sqrt{e}-4}{6\sqrt{e}-3}$.

        This implies that the total work area $W_2$ satisfies:
        \begin{align*}
            W_2 \geq \left(\frac{4\sqrt{e}-4}{6\sqrt{e}-3}\right) \cdot m_1.
        \end{align*}
        This strictly contradicts the upper bound on $W_2$ established in \Cref{obs:18}.
    \end{claimproof}

    We have now shown all the necessary properties to prove \Cref{lem:s2_eq_1}.
    Assume, for the sake of contradiction, that \cref{alg:RepairS2} returns an invalid schedule.
    Let $i$ be the machine with a total load of $\load(i)>\lambda \geq \frac{2\sqrt{e}}{2\sqrt{e}-1}$.
    By construction, this machine executes exactly two jobs, one from $S_1$ and one from $S_2$.
    We denote the processing time of the job from $S_2$ on that machine by
    $t$.
    Since this job must have been compressed at least once, we have $t\in (\lambda-1, \frac{4}{3}\frac{4\sqrt{e}-4}{6\sqrt{e}-3})$.

    Consider now any machine $i'\in \{i+1,\ldots,m_1\}$.
    Since the jobs in \(S_1\) are ordered in non-decreasing processing time, the job from \(S_1\) on machine \(i'\) has a processing time of at least \(\lambda-t\).
    By \Cref{obs:19}, we know that the job from \(S_2\) has a processing time of at least \(\frac{4}{10\sqrt{e}-5}\).
    Therefore:
    \begin{align}
        \label{eq:load-right}
        \load({i'}) \geq \lambda-t + \frac{4}{10\sqrt{e}-5} > \frac{\sqrt{e}+1}{2\sqrt{e}-1} &  & \forall i' \in \{i+1,\ldots,m_1\}
    \end{align}

    We make a similar argument for the machines to the left of $i$.

    Consider any machine $i'\in \{q+2,\ldots,i-1\}$:
    Since the jobs in \(S_2\) are ordered in non-increasing processing time, the job from \(S_2\) on machine \(i'\) has a processing time of at least \(t \geq \lambda-1\).
    Each job in \(S_1\) (with the exception of the job on machine \(q+1\)) has a processing time of at least \(\frac{\lambda}{2}\) by \Cref{cond:shelf1_height}.
    Therefore:
    \begin{align}
        \label{eq:load-left}
        \load({i'}) \geq \frac{\lambda}{2} + t \geq \frac{\lambda}{2} + (\lambda-1) = \frac{3\lambda}{2}-1 \geq \frac{\sqrt{e}+1}{2\sqrt{e}-1} &  & \forall i' \in \{q+2,\ldots,i-1\}
    \end{align}

    We now consider the leftout machines \(q+1\) and \(i\).
    Since the load on machine \(i\) is at least \(\lambda\), the combined load on these two machines is at least:
    \begin{align*}
        \lambda+(\lambda-1) & \geq 2\frac{2\sqrt{e}}{2\sqrt{e}-1}-1                            \\
                            & = 2\frac{2\sqrt{e}}{2\sqrt{e}-1}-\frac{2\sqrt{e}-1}{2\sqrt{e}-1} \\
                            & =\frac{2\sqrt{e}+1}{2\sqrt{e}-1}                                 \\
                            & =\frac{2\sqrt{e}+2}{2\sqrt{e}-1}-\frac{1}{2\sqrt{e}-1}           \\
                            & =2\frac{\sqrt{e}+1}{2\sqrt{e}-1}-\frac{1}{2\sqrt{e}-1}.
    \end{align*}
    Combining this with \cref{eq:load-right,eq:load-left}, we have that the total area of all machines from \(q+1\) to \(m_1\) is at least \(\frac{\sqrt{e}+1}{2\sqrt{e}-1}(m_1-q)-\frac{1}{2\sqrt{e}-1}\).
    Using that the load on the remaining \(q\) machines is at least \(\lambda - 1 = \frac{1}{2\sqrt{e}-1}\), we get a total area of at least:
    \begin{align*}
        \frac{\sqrt{e}+1}{2\sqrt{e}-1}(m_1-q)-\frac{1}{2\sqrt{e}-1}+\frac{1}{2\sqrt{e}-1}q.
    \end{align*}
    Since the total area of all jobs is at most \(m_1\) by \cref{cond:total_area}, we have:
    \begin{align*}
                 & \frac{\sqrt{e}+1}{2\sqrt{e}-1}(m_1-q)-\frac{1}{2\sqrt{e}-1}+\frac{1}{2\sqrt{e}-1}q          & < & m_1                           \\
        \implies & \frac{\sqrt{e}+1}{2\sqrt{e}-1}m_1-\frac{\sqrt{e}}{2\sqrt{e}-1}q  -\frac{1}{2\sqrt{e}-1}     & < & m_1                           \\
        \implies & \frac{\sqrt{e}+1}{2\sqrt{e}-1}m_1-\frac{2\sqrt{e}-1}{2\sqrt{e}-1}m_1 -\frac{1}{2\sqrt{e}-1} & < & \frac{\sqrt{e}}{2\sqrt{e}-1}q \\
        \implies & \frac{2-\sqrt{e}}{\sqrt{e}}m_1-\frac{1}{\sqrt{e}}                                           & < & q                             \\
        \implies & (\frac{2-\sqrt{e}}{\sqrt{e}}-\frac{1}{\sqrt{e}m_1})m_1                                      & < & q
    \end{align*}

    Note that $q \ge 1$ and the condition $q<\frac{1}{3}(1-\frac{1}{\sqrt{e}})m_1<0.132m_1$ from \Cref{obs:single_job_small_q} imply $m_1 > 7$. For $m_1 > 7$, it follows that $\frac{2-\sqrt{e}}{\sqrt{e}} - \frac{1}{\sqrt{e}m_1} > 0.137$, which reduces the inequality to $q > 0.137m_1$. This produces a direct contradiction. \qedhere
\end{proof}
\Cref{obs:single_job_large_q} and \Cref{obs:single_job_small_q} prove \Cref{lem:s2_eq_1}

\section{Worst Case Instance for \Cref{alg:assignment}}
\label{app:instance}
In this section, we give an instance so that \cref{alg:assignment} has an approximation ratio arbitrary close to $\frac{2\sqrt{e}}{2\sqrt{e}-1}$ showing that our analysis is tight.
\LowerBoundLemmaCmd*
\begin{figure}[t]
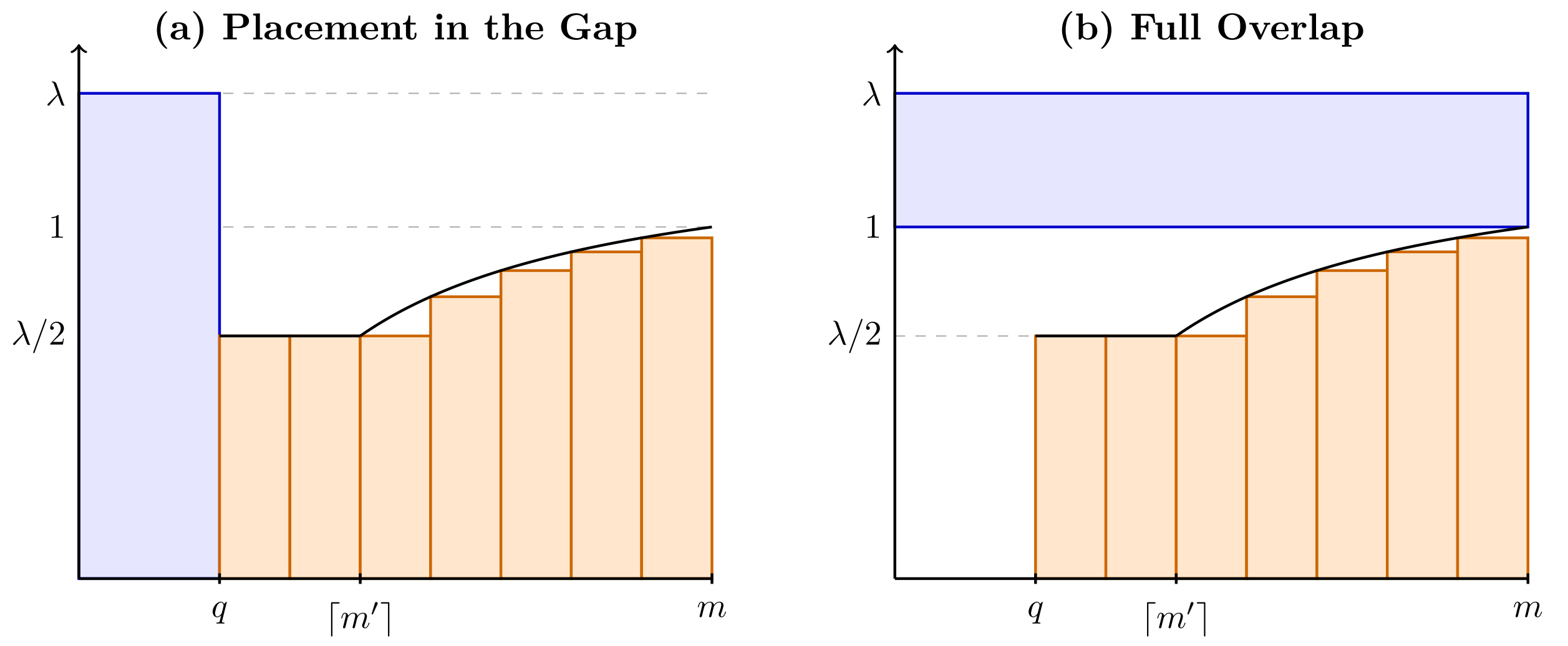

    \centering
    \includestandalone[width=\textwidth]{instance}
    \caption{The small jobs are shown in orange and the large job is depicted in blue.}
    \label{fig:instance}
\end{figure}

\begin{proof}
    Let \(\lambda = \frac{2\sqrt{e}}{2\sqrt{e}-1}\). We construct a family of instances parameterized by $\varepsilon$.
    We will show the ratio for some $\varepsilon'\leq \varepsilon$ for a sufficiently large number of machines \(m\) as chosen later.
    Let \(q = \frac{1}{2\sqrt{e}}m \) represent a specific machine index when rounded.
    We define an instance consisting of exactly \(m - \lfloor q\rfloor + 1\) jobs.
    One large job \(\job\) with an area of exactly \(\Aj = (\lambda - 1)m\) and \(m - \lfloor q\rfloor\) small jobs indexed by \(j \in \{\lfloor q\rfloor +1, \ldots, m\}\).

    We define a critical machine threshold \(m' = \frac{2\Aj}{\lambda} = \frac{2(\lambda-1)}{\lambda}m\).
    We carefully define the processing times of the small jobs so that if the big job is compressed and placed above them, the load on the rightmost machine is exactly \(\lambda\) (minus a small error).

    We define constant work area functions for each small job:
    \begin{equation}\label{eq:instance}
        w_j(\cdot)\coloneqq
        \begin{cases}
            \frac\lambda2\qquad                     & \text{if } j\in(q,\lceil m' \rceil)  \\
            \lambda - \frac{\Aj}{j} - \delta \qquad & \text{if } j\in [\lceil m'\rceil, m]
        \end{cases}
    \end{equation}
    where \(\delta = \frac{\varepsilon'+\frac\lambda2(q-\lfloor q\rfloor)}{m - \lceil m' \rceil+1}\) is a small error that ensures that the total area of the small jobs is exactly \(m-\Aj\) which will be shown in the next claim.
    Consequently, the total work area in the entire instance is exactly \(\Aj + (m-\Aj) = m\).
    Thus, an optimal schedule of makespan \(1\) can be obtained by assigning \(m\) machines to every job and executing them consecutively.

    \begin{claim}\label{claim:w1_area_instance}
        The total work area of the jobs defined by \cref{eq:instance} is $m-\Aj$.
    \end{claim}
    \begin{claimproof}
        The total work area of \cref{eq:instance} is:
        \begin{align}\label{eq:w1_def}
            W_1\coloneq \sum_{j=\lfloor q\rfloor +1}^m w_j(1) & =\frac\lambda2(m-\lfloor q\rfloor)+(m-\lceil m'\rceil +1)(\frac\lambda2-\delta)-\sum_{j=\lceil m'\rceil}^{m}\frac\Aj j                                         \\
                                                              & =\frac\lambda2(m-\lfloor q\rfloor )-(m-\lceil m'\rceil +1)\delta+\sum_{j=\lceil m'\rceil}^m\left[\frac\lambda2-\frac\Aj{j}\right]\nonumber                     \\
                                                              & =\frac\lambda2(m-\lfloor q\rfloor )-(\varepsilon'+\frac\lambda2(q-\lfloor q\rfloor))+\sum_{j=\lceil m'\rceil}^m\left[\frac\lambda2-\frac\Aj{j}\right]\nonumber \\
                                                              & =\frac\lambda2(m- q )-\varepsilon'+\sum_{j=\lceil m'\rceil}^m\left[\frac\lambda2-\frac\Aj{j}\right].\nonumber
        \end{align}
        Since $\sum_{i=\lceil m'\rceil}^m\left[\frac\lambda2-\frac\Aj i\right]$ is larger and converges to $\int_{m'}^m\frac\lambda2-\frac\Aj x\ dx$ (\cref{claim:sum_greater_int}), we can find an $\varepsilon'\leq 2\varepsilon$ and choose $m$ large enough so that $\frac{\frac\lambda2(q-\lfloor q\rfloor)}{m-\lceil m'\rceil+1}\leq \frac{\varepsilon}{2}$ and
        \begin{equation}\label{eq:int_sum_error}
            \sum_{i=\lceil m'\rceil}^m\left[\frac\lambda2-\frac\Aj i\right] = \int_{m'}^m\frac\lambda2-\frac\Aj x\ dx+\varepsilon'.
        \end{equation}
        Using now \cref{eq:w1_def,eq:int_sum_error}, we get
        \begin{equation}\label{eq:w1_int}
            W_1=\frac\lambda2(m-q)-\varepsilon'+\int_{m'}^m\frac\lambda2-\frac\Aj x\ dx+\varepsilon'.
        \end{equation}
        Further we have that
        \begin{align}
            (\lambda-1)m-\frac\lambda2q-\Aj\ln(\frac{m}{m'}) & =(\lambda-1)m-\frac\lambda2(\frac{1}{2\sqrt{e}}m)-(\lambda-1)\ln\left(\frac{m}{\frac{2(\lambda-1)}{\lambda}m}\right) \nonumber \\
                                                             & =(\lambda-1)m-\frac12(\lambda-1)m-\frac12(\lambda-1) = 0. \label{eq:f_value}
        \end{align}
        \cref{eq:f_value} implies now similar as in \cref{eq:amax} that
        \begin{align}\label{eq:int_equality}
                                 & (\lambda-1)m-\frac\lambda2q-\Aj\ln(\frac{m}{m'})= 0                                          \nonumber \\
            \Leftrightarrow\quad & (\lambda-1)m-\frac{\lambda}{2}\frac{2}{\lambda}\Aj-\frac\lambda2q-\Aj\ln(\frac{m}{m'})= -\Aj \nonumber \\
            \Leftrightarrow\quad & \frac\lambda2(m-q)+\frac{\lambda}{2}(m-\frac{2}{\lambda}\Aj)-\Aj\ln(\frac{m}{m'})= m-\Aj \nonumber     \\
            \Leftrightarrow\quad & \frac\lambda2(m-q)+\frac{\lambda}{2}(m-m')-\Aj\int_{m'}^{m}\frac{1}{x}\ dx= m-\Aj            \nonumber \\
            \Leftrightarrow\quad & \frac\lambda2(m-q)+\int_{m'}^{m}\frac{\lambda}{2}-\frac{\Aj}{x}\ dx= m-\Aj.
        \end{align}
        Combining now \cref{eq:w1_int,eq:int_equality} results in
        \begin{align*}
            W_1=m-\Aj. \tag*{\qedhere}
        \end{align*}
    \end{claimproof}

    To show the claimed approximation ratio, we now have to show that the algorithm returns a schedule with makespan \(\lambda = \frac{2\sqrt{e}}{2\sqrt{e}-1}\).

    First, observe that due to the constant work functions any partition \(C_1 \sqcup C_2 \sqcup C_3\) satisfies the properties in \cref{thm:base_properties}.
    Thus, \cref{alg:jansen_ohnesorge} places small jobs in \(S_0\sqcup S_1\) and the single large job \(\job\) is placed in \(S_2\).
    The algorithm will then assign the small jobs on one machine each in non-decreasing order of processing time to machines \(\lfloor q\rfloor +1, \ldots, m\).
    Then, since \(|S_2|=1\) \Cref{alg:assignment} will attempt to place the big job \(\job\) on machines \(1, \ldots, i\) for some \(i \in [m]\). We examine the resulting makespan for all possible choices of \(i\), two special cases are shown in \cref{fig:instance}:

    \noindent\textbf{Case 1 (Placement in the gap):}
    If \(\job\) is compressed to exactly \(i = \lfloor q\rfloor\) machines, it fits into the empty gap without overlapping any of the small jobs.
    Its processing time is:
    \begin{align*}
        \frac{\Aj}{\lfloor q\rfloor} = \frac{(\lambda - 1)m}{\lfloor\frac{1}{2\sqrt{e}}m\rfloor}\geq \frac{(\lambda - 1)m}{\frac{1}{2\sqrt{e}}m} = 2\sqrt{e}(\lambda - 1) = 2\sqrt{e}\left(\frac{2\sqrt{e}}{2\sqrt{e}-1}-\frac{2\sqrt{e}-1}{2\sqrt{e}-1}\right)= \lambda.
    \end{align*}

    \noindent\textbf{Case 2 (Partial overlap):}
    If \(\job\) is compressed onto \((q, \lceil m' \rceil)\) machines, it overlaps with the first block of small jobs.
    The total load on machine \(i\) is the small jobs processing time plus the processing time of the big job on \(i\) machines. By definition of \(m'\) in \cref{eq:m'_condition}, we get:
    \begin{align*}
        \frac{\lambda}{2} + \frac{\Aj}{i} \geq \frac{\lambda}{2} + \frac{\Aj}{m'} > \frac{\lambda}{2} + \frac{\lambda}{2} = \lambda.
    \end{align*}

    \noindent\textbf{Case 3 (Full overlap):}
    If \(\job\) is compressed onto \([\lceil m' \rceil, m]\) machines, it overlaps with the second block of small jobs.
    The total load on machine \(i\) is again the small jobs processing time plus the processing time of the big job:
    \begin{align*}
        t_j(i) + \frac{\Aj}{i} = \left(\lambda - \frac{\Aj}{i} - \delta \right) + \frac{\Aj}{i} = \lambda - \delta.
    \end{align*}

    Since \(\delta = \frac{\varepsilon'+\frac\lambda2(q-\lfloor q\rfloor)}{m - \lceil m' \rceil+1}\leq\frac{\frac{\varepsilon}{2}+\frac{\varepsilon}{2}}{m-\lceil m'\rceil+1}\leq \varepsilon\), the resulting makespan is at least $\lambda-\varepsilon$.
    \qedhere
\end{proof}

\section{Edge Cases}
\label{sec:edge_case}
In this section we prove the following claim:
\edgecases*
We will archive this by proving each individual claim.
Note that the $q+1<\lceil m'\rceil$ implies $q+2\leq \lceil m'\rceil -1$ since $q$ is an integer, so in the following the prove the claim with this condition instead.

\begin{observation}
    \label{lem:load_sum_lower_bound}
    If $q+2\leq \lceil m'\rceil -1$, then we can transform the schedule given by \cref{rem:schedule} so that the following property holds as well:
    \begin{equation*}
        \displaystyle\sum_{i=q+1}^{\lceil m'\rceil-1}\load(i)\geq \frac\lambda2(\lceil m'\rceil -q -1)
    \end{equation*}
\end{observation}
\begin{proof}
    In this case we assume that the job $j^1$ from \cref{cond:shelf1_height} with processing time at most $\frac\lambda2$ exists.
    If this job would not exits, any job in $S_1$ would have processing time of at least $\frac\lambda2$ which directly implies $\sum_{i=q+1}^{\lceil m'\rceil -1}\load(i)\geq \frac\lambda2(\lceil m' \rceil -q -1)$.
    Note that the case $q+2\leq \lceil m'\rceil -1$ is chosen so that the sum consists of at least two terms which guarantees that the sum contains the indices $q+1$ and $q+2$.

    \noindent\textbf{Case 1 (\(\load(q+1)+\load(q+2)\leq \lambda\)):}
    Since $j^1$ is the job with lowest processing time, and the schedule is increasing in load, $j^1$ is scheduled on machine $q+1$.
    We can reschedule $j^1$ onto machine $q+2$.
    Note that this changes $q$.
    Since the combined load is at most $\lambda$, we do not increase the makespan beyond $\lambda$.
    We will then change the schedule and make it non-contiguous by sorting the machine $q+2$ into the right position such that the load is again non-decreasing.
    This simplifies the following analysis, since the schedule has now the same properties as before (except for contiguousness).
    If $\load(q+1)+\load(q+2)>1$, we decrease $m_1$ by 1 since this machine is then part of $S_0$.
    After placing the job in $S_2$ with \cref{alg:assignment} we can make this schedule contiguous again by moving the separated job to the left.
    This transformation is visualized in \cref{fig:reschedule}.
    The algorithm is now extended by using steps (a)-(c) as preprocessing and step (e) as postprocessing.
    Since after step (c) every machine between $q+1$ and $m_1$ has load of at least $\frac\lambda2$, the statement holds.
    \input{edge_case}

    \noindent\textbf{Case 2 (\(\load(q+1)+\load(q+2)>\lambda\)):}
    In this case we can bound the sum by $\sum_{q+1}^{\lceil m'\rceil -1}\load(i)=\load(q+1)+\load(q+2)+\sum_{i=q+3}^{\lceil m'\rceil -1}\load(i)\\ >\lambda + \frac\lambda2\left(\left(\lceil (m'\rceil-1\right)-(q+3)+1 \right)=\frac\lambda2(\lceil m' \rceil -q-1)$ by \ref{cond:shelf1_height}.
\end{proof}

\begin{restatable}{observation}{edgecase}
    \label{lem:eq_edge_case}
    Let $\lambda\in[\frac{10}7,\frac32]$ and $q+1=\lceil m'\rceil -1$, then \cref{alg:assignment} has ratio $\lambda$.
\end{restatable}

\begin{proof}
    We can assume $\Aj>\lambda q$ since otherwise $\Aj\leq \lambda q$ would hold, which means that we can compress $\job$ onto $q$ machines with processing time at most $\lambda$.
    This condition is equivalent to $\frac{\Aj}{\lambda}>q$.
    Which implies $\frac{\Aj}{\lambda}+1>q+1=\lceil m'\rceil -1\geq m'-1=\frac{2\Aj}{\lambda}-1$ by our initial assumption.
    This can be transformed to \begin{equation}\label{eq:0}2\geq \frac{\Aj}{\lambda}\Leftrightarrow 2\lambda \geq \Aj\end{equation} which implies that we can compress $\job$ onto $2$ machines with processing time at most $\lambda$.
    If $q\geq 2$ holds, then we would do this and be done.
    The remaining case is then $q=1$ (since the ratio for the case $q=0$ has been shown in \cite{ohnesorge}).
    To proof that the ratio holds for all $m_1$, we will state some properties which must hold if \cref{alg:assignment} produces a schedule with makespan greater $\lambda$ and show that they are contradicting.
    Define $W_1\coloneq \sum_{j\in S_1}w_j(\gamma_j(1))$.
    \begin{align}
        \label{eq:1} \load(i)+\frac{\Aj}{i} & > \lambda \quad \text{for all } i\in \{2,\ldots,m_1\} &  & \text{since $\job$ is not compressed onto $i$ machines.} \\
        \label{eq:2} \Aj                    & > \lambda                                             &  & \text{by \Cref{cond:shelf2_area}}                        \\
        \label{eq:3} W_1+\Aj                & \leq m_1                                              &  & \text{by \Cref{cond:total_area}}.
    \end{align}

    \noindent\textbf{Case 1 (\(m_1 = 2\)):}
    We have:
    \begin{align*}
        W_1+\Aj=\load(2)+\Aj \overset{\cref{eq:1}}{>}\lambda -\frac{\Aj}2+\Aj=\lambda+\frac\Aj2\overset{\cref{eq:2}}{>}\frac32\lambda >2=m_1
    \end{align*}
    Which is a contradiction to \cref{eq:3}.

    \noindent\textbf{Case 2 (\(m_1 = 3\)):}
    This results in a contradiction as well:
    $W_1+\Aj=\load(2)+\load(3)+\Aj\overset{\cref{eq:1}}{>}\lambda-\frac{\Aj}2+\lambda -\frac\Aj3+\Aj=2\lambda +\frac16\Aj\overset{\cref{eq:2}}{>}2\lambda+\frac16\lambda>3=m_1$.

    \noindent\textbf{Case 3 (\(m_1 = 4\)):}
    In this case we obtain a contradiction by:
    $W_1+\Aj=\load(2)+\load(3)+\load(4)+\Aj>3\lambda -\frac\Aj2-\frac\Aj3-\frac\Aj4+\Aj=3\lambda-\frac1{12}\Aj\overset{\cref{eq:0}}{>}3\lambda-\frac2{12}\lambda>4=m_1$.

    \noindent\textbf{Case 4 (\(m_1 \geq 5\)):}
    In this remaining case we can use that $\Aj$ compressed onto $m_1$ machines must have processing time greater than $\lambda-1$ otherwise we could place it there and be done.
    This will lead us similarly to a contradiction, showing that we need more area than we maximal have by \ref{cond:total_area}.
    \begin{align*}
        W_1+\Aj & \overset{\ref{cond:shelf1_area}}>\frac{\lambda}{2}(m_1-1)+\Aj & \overset{\ref{eq:work_job_lowerbound}}> & \frac\lambda2(m_1-1)+(\lambda-1)m_1 \\
                & = \frac{\lambda}2m_1-\frac\lambda2+\lambda m_1-m_1            & =                                       & \lambda(\frac32m_1-\frac12)-m_1     \\
                & \geq \frac{10}7(\frac32m_1-\frac12)-m_1                       & =                                       & \frac{8}7m_1-\frac57                \\
                & =m_1+(\frac17m_1-\frac57)                                     & \overset{m_1\geq 5}\geq                 & m_1
    \end{align*}
    Again a contradiction to \cref{eq:3}.
\end{proof}
\Cref{lem:load_sum_lower_bound,lem:eq_edge_case} now prove \cref{prop:lem_sufficient_condition}.

\section{Analysis Proofs}
In this section we prove the following claim:
\anaproperties*
We will archive this by proving each claim individually in the following.

\label{sec:ana_proofs}

\let\oldtheclaim\theclaim
\renewcommand{\theclaim}{2.3.1}
\begin{claim}
    $\displaystyle\sum_{i=\lceil{m'}\rceil}^{m_1}\left[\frac\lambda2-\frac{\Aj}i\right]>\int_{m'}^{m_1}\frac\lambda2-\frac{\Aj}x\ dx $
\end{claim}
\let\theclaim\oldtheclaim
\addtocounter{claim}{-1}
\begin{claimproof}
    Define $h(i)\coloneq \frac\lambda2-\frac{\Aj}{i}$ and $k\coloneq \lceil m'\rceil$.

    Note that by \cref{eq:m'_condition}, $h(i)\geq 0$ holds, and for every $i\in [1,m_1]$, the following holds, since \(h(i)\) is strictly monotone:
    \begin{align*}
        \int_{i-1}^{i}h(x)\ dx< \int_{i-1}^{{i}}h(i)\ dx= h(i)
    \end{align*}
    This implies:
    \begin{equation*}
        \int_k^{m_1}h(x)\ dx=\sum_{i=k+1}^{m_1}\int_{i-1}^i h(x)\ dx\leq \sum_{i=k+1}^{m_1}h(i)
    \end{equation*}
    and similarly for the rounded part:
    \begin{equation*}
        \int_{m'}^{k}h(x)\ dx \leq \int_{m'}^kh(k)\ dx =(k-m')h(k)<h(k)
    \end{equation*}
    Combined, we have:
    \begin{equation*}
        \int_{m'}^{m_1}h(x)\ dx=\int_{m'}^kh(x)\ dx +\int_k^{m_1}h(x)\ dx < h(k)+\sum_{i=k+1}^{m_1}h(i)=\sum_{i=\lceil m'\rceil}^{m_1}h(i). \qedhere
    \end{equation*}

\end{claimproof}

\let\oldtheclaim\theclaim
\renewcommand{\theclaim}{2.3.2}
\begin{claim}
    \label{lem:f_monoton}
    If $\lambda \in [\frac{4}{3},\frac{3}{2}]$ and $q\leq \qc$, then $f(q)=(\lambda-1)m_1-\frac\lambda2q-\Aj\ln\left(\frac{m_1}{m'}\right)$ is monotonically decreasing in $q$.
\end{claim}
\let\theclaim\oldtheclaim
\addtocounter{claim}{-1}
\begin{claimproof}
    Considering the first derivative of $f$ with respect to $q$, we have:
    \begin{align*}
        \frac{\partial f}{\partial q} = -\frac{\lambda}{2}.
    \end{align*}
    Given the bounded interval $\lambda \in [\frac{4}{3},\frac{3}{2}]$, it trivially holds that $-\frac{\lambda}{2} \leq -\frac{2}{3} < 0$.
\end{claimproof}

\let\oldtheclaim\theclaim
\renewcommand{\theclaim}{2.3.3}
\begin{claim}
    \label{lem:q_mon}
    If $\lambda\in [1,\frac{3}{2}]$, then $\qc= \frac{(1-c)(\lambda -1)m_1+c(1-\frac\lambda2)m_1}{(1-\frac{c}2)\lambda}$ is monotonically increasing in $c$.
\end{claim}
\let\theclaim\oldtheclaim
\addtocounter{claim}{-1}

\begin{claimproof}
    We will calculate the derivative of $\qc$ again.
    Factor out the constant \(\frac{m_1}{\lambda}\) and simplify the numerator $u(c)$:
    \begin{align*}
        u(c) & = (1-c)(\lambda - 1) + c\left(1 - \frac{\lambda}{2}\right) \\
             & = \lambda - 1 + 2c - \frac{3}{2}c\lambda
    \end{align*}
    Define $v(c)$ as the denominator and calculate derivatives:
    \begin{align*}
        u'(c) & = 2 - \frac{3}{2}\lambda                        \\
        v(c)  & = 1 - \frac{c}{2} \implies v'(c) = -\frac{1}{2}
    \end{align*}
    Apply the quotient rule to find the new numerator $u'v - uv'$:
    \begin{align*}
        u'v - uv' & = \left(2 - \frac{3}{2}\lambda\right)\left(1 - \frac{c}{2}\right) - \left(\lambda - 1 + 2c - \frac{3}{2}c\lambda\right)\left(-\frac{1}{2}\right) \\
                  & = \left( 2 - c - \frac{3}{2}\lambda + \frac{3}{4}c\lambda \right) + \left( \frac{1}{2}\lambda - \frac{1}{2} + c - \frac{3}{4}c\lambda \right)    \\
                  & = \frac{3}{2} - \lambda = \frac{3 - 2\lambda}{2}.
    \end{align*}
    Combine all parts with the factored constants $\frac{m_1}{\lambda}$ to get the final derivative:
    \begin{align*}
        \frac{\partial \qc}{\partial c} & = \frac{m_1}{\lambda} \cdot \frac{ \frac{3 - 2\lambda}{2} }{ \left(1 - \frac{c}{2}\right)^2 } \\
                                        & = \frac{m_1(3 - 2\lambda)}{2\lambda\left(1 - \frac{c}{2}\right)^2}.
    \end{align*}
    We now have that:
    \begin{align*}
        \frac{\partial \qc}{\partial c}=\frac{m_1(3-2\lambda)}{2\lambda(1-\frac{c}{2})^2}\stackrel{!}\geq 0\Leftrightarrow 3-2\lambda\geq 0\Leftrightarrow \frac{3}{2}\geq \lambda. \tag*{\qedhere}
    \end{align*}

\end{claimproof}

\let\oldtheclaim\theclaim
\renewcommand{\theclaim}{2.3.4}
\begin{claim}
    \label{lem:g_mon}
    If $\lambda \geq \frac{2\sqrt{e}}{2\sqrt{e}-1}$, then $g(\qc)=(\lambda-1)m_1-\frac{\lambda}{2}\qc-\lambda \qc\ln\frac{m_1\lambda}{2\lambda \qc}$ is monotonically increasing in $\qc$.
\end{claim}
\let\theclaim\oldtheclaim
\addtocounter{claim}{-1}

\begin{claimproof}
    First, we calculate the derivative of $g$.
    Cancel $\lambda$ inside the logarithm to simplify:
    \begin{align*}
        \ln \left( \frac{m_1 \lambda}{2\lambda \qc} \right) & = \ln \left( \frac{m_1}{2\qc} \right)
    \end{align*}

    Differentiate term by term with respect to $\qc$:
    \begin{align*}
        \frac{\partial g}{\partial \qc} & = \frac{\partial}{\partial \qc} \left[ (\lambda - 1)m_1 - \frac{\lambda}{2}\qc - \lambda \qc \ln \left( \frac{m_1}{2\qc} \right) \right] \\
                                        & = 0 - \frac{\lambda}{2} - \frac{\partial}{\partial \qc} \left[ \lambda \qc \ln \left( \frac{m_1}{2\qc} \right) \right]
    \end{align*}

    Apply the product rule $(uv)' = u'v + uv'$ to the last term:
    \begin{align*}
        \frac{\partial}{\partial \qc} \left[ \lambda \qc \ln \left( \frac{m_1}{2\qc} \right) \right] & = \lambda \ln \left( \frac{m_1}{2\qc} \right) + \lambda \qc \left( -\frac{1}{\qc} \right) \\
                                                                                                     & = \lambda \ln \left( \frac{m_1}{2\qc} \right) - \lambda
    \end{align*}

    Substitute back, combine constants and factor out $\frac{\lambda}{2}$:
    \begin{align*}
        \frac{\partial g}{\partial \qc} & = -\frac{\lambda}{2} - \left( \lambda \ln \left( \frac{m_1}{2\qc} \right) - \lambda \right) \\
                                        & = \frac{\lambda}{2} - \lambda \ln \left( \frac{m_1}{2\qc} \right)                           \\
                                        & = \frac{\lambda}{2} \left( 1 - 2 \ln \left( \frac{m_1}{2\qc} \right) \right)
    \end{align*}

    We now have:
    \begin{align*}
        \frac{\partial g(c)}{\partial \qc}= & \frac{\lambda}{2}\left(1-2\ln \frac{m_1}{2\qc}\right)\stackrel{!}{\geq}0 \\
        \Leftrightarrow\quad                & \frac{1}{2}\geq \ln \frac{m_1}{2 \qc}                                    \\
        \Leftrightarrow\quad                & \sqrt{e}\geq \frac{m_1}{2\qc}                                            \\
        \Leftrightarrow\quad                & \qc\geq \frac{m_1}{2\sqrt{e}}
    \end{align*}
    And:
    \begin{align*}
                        & \lambda\geq \frac{2\sqrt{e}}{2\sqrt{e}-1}=\frac{1}{1-\frac{1}{2\sqrt{e}}} \\
        \Leftrightarrow & \lambda(1-\frac{1}{2\sqrt{e}})\geq 1                                      \\
        \Leftrightarrow & \lambda-1\geq \lambda\frac{1}{2\sqrt{e}}                                  \\
        \Leftrightarrow & \frac{\lambda-1}{\lambda}m_1\geq \frac{m_1}{2\sqrt{e}} \tag*{\qedhere}
    \end{align*}
\end{claimproof}
\section{The Full Algorithm}
\label{sec:assign_partitions}
In this section we restate the full algorithm from~\cite{ohnesorge} for completeness and clarity.

The partition \(C_1 \sqcup C_2 \sqcup C_3 = J_B\) is found by solving the \ac{mckp} as formulated in \Cref{lem:main-partition}.
The high-level overview of this algorithm is given in \Cref{alg:jansen_ohnesorge}.
Specifically, line 2 and 3 of \Cref{alg:jansen_ohnesorge} are implemented in \Cref{alg:assign_partitions}.

\begin{algorithm}[ht]
    \caption{Algorithm Assign Partitions}
    \label{alg:assign_partitions}
    \begin{algorithmic}[1]
        \Require Partition $C_1, C_2, C_3$ as in \cref{lem:main-partition}, $\lambda\in [\frac{2\sqrt{e}}{2\sqrt{e}-1},\frac32]$ and makespan guess $d=1$.

        \State For each $j \in C_1 \sqcup C_2$, set $\alpha_j \gets \gamma_j(\lambda)$; move $j$ to $S_0$ if $t_j(\alpha_j) > 1$, else to $S_1$
        \State Schedule $S_0 \sqcup S_1$ on $\sum_{j\in S_0\sqcup S_1} \alpha_j$ machines in non-decreasing order of processing times \\\Comment{May use more than $m$ machines.}

        \While{there exist distinct machines $i_1, i_2$ assigned only jobs with $\alpha_j=1$ and $\mathrm{load}(i_1) \le \mathrm{load}(i_2) \le \frac{\lambda}{2}$}
        \State Move all jobs from $i_2$ to $i_1$
        \EndWhile

        \If{there exists a machine $i$ assigned only jobs with $\alpha_j=1$ such that $\mathrm{load}(i) \le \frac{\lambda}{2}$ and $\mathrm{load}(i) + \mathrm{load}(m_1) \le \lambda$}
        \State Move all jobs from $i$ to $m_1$
        \EndIf

        \While{there exist distinct machine triplets $\{i_1, i_1+1, i_1+2\}$ and $\{i_2, i_2+1, i_2+2\}$ assigned only jobs with $\alpha_j=3$ and $\mathrm{load}(i_1) \le \mathrm{load}(i_2) \le \frac{\lambda}{2}$}
        \State Move all jobs from $\{i_2, i_2+1, i_2+2\}$ to $\{i_1, i_1+1, i_1+2\}$
        \EndWhile
        \Comment{Now at most $m$ machines are used.}
        \State Denote by $q\coloneq m-\sum_{i, load(i)>0}1$ the number of unused machines
        \For{each job $j \in C_3$}
        \State Set $\alpha_j \gets \gamma_j(\lambda - 1)$
        \If{$w_j(\alpha_j) \le \lambda q$ \textbf{or} $\alpha_j \le q$}
        \State Add $j$ to $C_1\sqcup C_2$ and re-apply previous steps (lines 1--13)
        \Else
        \State Add $j$ to $S_2$.
        \EndIf
        \EndFor
        \State Remove unused machines so that only $m$ machines in total remain
        \State Sort the machines (contiguously) by load non-decreasingly
    \end{algorithmic}
\end{algorithm}
\end{document}

%% file: instance.tex
\begin{tikzpicture}[
        x=0.72cm,
        y=3.6cm,
        every node/.style={font=\small},
        axis/.style={black, thick},
        tick/.style={black, thick},
        curve/.style={black, thick},
        smalljob/.style={
                fill=orange!20,
                draw=orange!80!black,
                thick
            },
        bigjob/.style={
                fill=blue!10,
                draw=blue!80!black,
                thick
            },
        guide/.style={
                gray!60,
                dashed,
                thin
            }
    ]

    \def\lam{1.38}
    \pgfmathsetmacro{\halflam}{\lam/2}

    \def\qpos{2}
    \def\mppos{4}
    \def\mpos{9}

    %
    %
    %
    \pgfmathsetmacro{\Wtwo}{
        (1-\lam/2)/(1/\mppos - 1/\mpos)
    }

    \pgfmathsetmacro{\deltaVal}{
        \lam - 1 - \Wtwo/\mpos
    }

    \begin{scope}[shift={(0,0)}]

        \node[font=\bfseries]
        at ({\mpos/2},{\lam+0.18})
        {(a) Placement in the Gap};

        \draw[guide] (0,\lam) -- (\mpos,\lam);
        \draw[guide] (0,1) -- (\mpos,1);
        \draw[guide] (0,\halflam) -- (\mppos,\halflam);

        %
        \draw[bigjob]
        (0,0) rectangle (\qpos,\lam);

        \foreach \a/\b in {
                2/3,
                3/4
            }{
                \draw[smalljob]
                (\a,0) rectangle (\b,\halflam);
            }

        %
        \foreach \a/\b in {
                4/5,
                5/6,
                6/7,
                7/8,
                8/9
            }{
                \pgfmathsetmacro{\h}{
                    \lam - \Wtwo/\a - \deltaVal
                }

                \draw[smalljob]
                (\a,0) rectangle (\b,\h);
            }

        \draw[curve]
        (\qpos,\halflam)
        --
        (\mppos,\halflam);

        \draw[
            curve,
            domain=\mppos:\mpos,
            samples=120,
            smooth
        ]
        plot (\x,{
            \lam - \Wtwo/\x - \deltaVal
        });

        \draw[axis,->]
        (0,0) -- (0,\lam+0.14);

        \draw[axis]
        (0,0) -- (\mpos,0);

        \draw[tick]
        (\qpos,0.015) -- (\qpos,-0.015);

        \draw[tick]
        (\mppos,0.015) -- (\mppos,-0.015);

        \draw[tick]
        (\mpos,0.015) -- (\mpos,-0.015);

        \node[below=3pt]
        at (\qpos,0) {$q$};

        \node[below=3pt]
        at (\mppos,0) {$\lceil m' \rceil$};

        \node[below=3pt]
        at (\mpos,0) {$m$};

        \draw[tick]
        (0.015,\lam) -- (-0.015,\lam);

        \draw[tick]
        (0.015,1) -- (-0.015,1);

        \draw[tick]
        (0.015,\halflam) -- (-0.015,\halflam);

        \node[left]
        at (0,\lam) {$\lambda$};

        \node[left]
        at (0,1) {$1$};

        \node[left]
        at (0,\halflam) {$\lambda/2$};

    \end{scope}

    \begin{scope}[shift={(11.6,0)}]

        \node[font=\bfseries]
        at ({\mpos/2},{\lam+0.18})
        {(b) Full Overlap};

        \draw[guide] (0,\lam) -- (\mpos,\lam);
        \draw[guide] (0,1) -- (\mpos,1);
        \draw[guide] (0,\halflam) -- (\mppos,\halflam);

        \draw[bigjob]
        (0,1) rectangle (\mpos,\lam);

        \foreach \a/\b in {
                2/3,
                3/4
            }{
                \draw[smalljob]
                (\a,0) rectangle (\b,\halflam);
            }

        \foreach \a/\b in {
                4/5,
                5/6,
                6/7,
                7/8,
                8/9
            }{
                \pgfmathsetmacro{\h}{
                    \lam - \Wtwo/\a - \deltaVal
                }

                \draw[smalljob]
                (\a,0) rectangle (\b,\h);
            }

        \draw[curve]
        (\qpos,\halflam)
        --
        (\mppos,\halflam);

        \draw[
            curve,
            domain=\mppos:\mpos,
            samples=120,
            smooth
        ]
        plot (\x,{
            \lam - \Wtwo/\x - \deltaVal
        });

        \draw[axis,->]
        (0,0) -- (0,\lam+0.14);

        \draw[axis]
        (0,0) -- (\mpos,0);

        \draw[tick]
        (\qpos,0.015) -- (\qpos,-0.015);

        \draw[tick]
        (\mppos,0.015) -- (\mppos,-0.015);

        \draw[tick]
        (\mpos,0.015) -- (\mpos,-0.015);

        \node[below=3pt]
        at (\qpos,0) {$q$};

        \node[below=3pt]
        at (\mppos,0) {$\lceil m' \rceil$};

        \node[below=3pt]
        at (\mpos,0) {$m$};

        \draw[tick]
        (0.015,\lam) -- (-0.015,\lam);

        \draw[tick]
        (0.015,1) -- (-0.015,1);

        \draw[tick]
        (0.015,\halflam) -- (-0.015,\halflam);

        \node[left]
        at (0,\lam) {$\lambda$};

        \node[left]
        at (0,1) {$1$};

        \node[left]
        at (0,\halflam) {$\lambda/2$};

    \end{scope}

\end{tikzpicture}

%% file: edge_case.tex
\begin{figure}[htbp]
\centering
\begin{subfigure}[b]{0.19\textwidth}
\centering
\begin{tikzpicture}[xscale=0.26, yscale=0.6, font=\sffamily]

\def\m{8} \def\opt{2} \def\mylambda{1.41}
\pgfmathsetmacro{\lamopt}{\mylambda * \opt}

\draw[thick] (0, 0) -- (\m, 0);
\draw[thick] (0, 0) -- (0, \lamopt);

\draw[thick] (\m, 0.1) -- (\m, -0.1) node[below] {$m_1$};
\draw[thick] (0.1, \lamopt) -- (-0.1, \lamopt) node[left] {$\lambda$};

\draw[dashed, gray!60] (0, \lamopt) -- (\m, \lamopt);

\filldraw[fill=orange!20, draw=orange!80!black, thick] (7, 0) rectangle (8, 1.85);
\filldraw[fill=orange!20, draw=orange!80!black, thick] (6, 0) rectangle (7, 1.75);
\filldraw[fill=orange!20, draw=orange!80!black, thick] (5, 0) rectangle (6, 1.65);
\filldraw[fill=orange!20, draw=orange!80!black, thick] (4, 0) rectangle (5, 1.55);
\filldraw[fill=orange!50, draw=orange!80!black, thick] (3, 0) rectangle (4, 1.4);
\filldraw[fill=orange!50, draw=orange!80!black, thick] (2.5, 0) rectangle (3, 1.2);

\draw[|-|, thick] (0, -0.55) -- (2.5, -0.55) node[midway, below=2pt] {$q$};

\end{tikzpicture}
\caption{Initial state.}
\end{subfigure}%
\hfill%
\begin{subfigure}[b]{0.19\textwidth}
\centering
\begin{tikzpicture}[xscale=0.26, yscale=0.6, font=\sffamily]

\def\m{8} \def\opt{2} \def\mylambda{1.41}
\pgfmathsetmacro{\lamopt}{\mylambda * \opt}

\draw[thick] (0, 0) -- (\m, 0);
\draw[thick] (0, 0) -- (0, \lamopt);

\draw[thick] (\m, 0.1) -- (\m, -0.1) node[below] {$m_1$};
\draw[thick] (0.1, \lamopt) -- (-0.1, \lamopt) node[left] {$\lambda$};

\draw[dashed, gray!60] (0, \lamopt) -- (\m, \lamopt);

\filldraw[fill=orange!20, draw=orange!80!black, thick] (7, 0) rectangle (8, 1.85);
\filldraw[fill=orange!20, draw=orange!80!black, thick] (6, 0) rectangle (7, 1.75);
\filldraw[fill=orange!20, draw=orange!80!black, thick] (5, 0) rectangle (6, 1.65);
\filldraw[fill=orange!20, draw=orange!80!black, thick] (4, 0) rectangle (5, 1.55);
\filldraw[fill=orange!50, draw=orange!80!black, thick] (3, 0) rectangle (4, 1.4);
\filldraw[fill=orange!50, draw=orange!80!black, thick] (3, 1.4) rectangle (3.5, 2.6);

\draw[|-|, thick] (0, -0.55) -- (3, -0.55) node[midway, below=2pt] {$q$};

\end{tikzpicture}
\caption{Reschedule $j^*$.}
\end{subfigure}%
\hfill%
\begin{subfigure}[b]{0.19\textwidth}
\centering
\begin{tikzpicture}[xscale=0.26, yscale=0.6, font=\sffamily]

\def\m{8} \def\opt{2} \def\mylambda{1.41}
\pgfmathsetmacro{\lamopt}{\mylambda * \opt}

\draw[thick] (0, 0) -- (\m, 0);
\draw[thick] (0, 0) -- (0, \lamopt);

\draw[thick] (\m, 0.1) -- (\m, -0.1) node[below] {$m_1$};
\draw[thick] (0.1, \lamopt) -- (-0.1, \lamopt) node[left] {$\lambda$};

\draw[dashed, gray!60] (0, \lamopt) -- (\m, \lamopt);

\filldraw[fill=orange!50, draw=orange!80!black, thick] (3, 0) rectangle (3.5, 1.4);

\filldraw[fill=orange!20, draw=orange!80!black, thick] (3.5, 0) rectangle (4.5, 1.55);
\filldraw[fill=orange!20, draw=orange!80!black, thick] (4.5, 0) rectangle (5.5, 1.65);
\filldraw[fill=orange!20, draw=orange!80!black, thick] (5.5, 0) rectangle (6.5, 1.75);
\filldraw[fill=orange!20, draw=orange!80!black, thick] (6.5, 0) rectangle (7.5, 1.85);

\filldraw[fill=orange!50, draw=orange!80!black, thick] (7.5, 0) rectangle (8, 1.4);
\filldraw[fill=orange!50, draw=orange!80!black, thick] (7.5, 1.4) rectangle (8, 2.6);

\draw[|-|, thick] (0, -0.55) -- (3, -0.55) node[midway, below=2pt] {$q$};

\end{tikzpicture}
\caption{Reorder.}
\end{subfigure}%
\hfill%
\begin{subfigure}[b]{0.19\textwidth}
\centering
\begin{tikzpicture}[xscale=0.26, yscale=0.6, font=\sffamily]

\def\m{8} \def\opt{2} \def\mylambda{1.41}
\pgfmathsetmacro{\lamopt}{\mylambda * \opt}

\draw[thick] (0, 0) -- (\m, 0);
\draw[thick] (0, 0) -- (0, \lamopt);

\draw[thick] (\m, 0.1) -- (\m, -0.1) node[below] {$m_1$};
\draw[thick] (0.1, \lamopt) -- (-0.1, \lamopt) node[left] {$\lambda$};

\draw[dashed, gray!60] (0, \lamopt) -- (\m, \lamopt);

\filldraw[fill=blue!10, draw=blue!80!black, thick] (0, 1.75) rectangle (6.5, \lamopt);

\filldraw[fill=orange!50, draw=orange!80!black, thick] (3, 0) rectangle (3.5, 1.4);
\filldraw[fill=orange!20, draw=orange!80!black, thick] (3.5, 0) rectangle (4.5, 1.55);
\filldraw[fill=orange!20, draw=orange!80!black, thick] (4.5, 0) rectangle (5.5, 1.65);
\filldraw[fill=orange!20, draw=orange!80!black, thick] (5.5, 0) rectangle (6.5, 1.75);
\filldraw[fill=orange!20, draw=orange!80!black, thick] (6.5, 0) rectangle (7.5, 1.85);
\filldraw[fill=orange!50, draw=orange!80!black, thick] (7.5, 0) rectangle (8, 1.4);
\filldraw[fill=orange!50, draw=orange!80!black, thick] (7.5, 1.4) rectangle (8, 2.6);

\draw[|-|, thick] (0, -0.55) -- (3, -0.55) node[midway, below=2pt] {$q$};

\end{tikzpicture}
\caption{\Cref{alg:assignment}.}
\end{subfigure}%
\hfill%
\begin{subfigure}[b]{0.19\textwidth}
\centering
\begin{tikzpicture}[xscale=0.26, yscale=0.6, font=\sffamily]

\def\m{8} \def\opt{2} \def\mylambda{1.41}
\pgfmathsetmacro{\lamopt}{\mylambda * \opt}

\draw[thick] (0, 0) -- (\m, 0);
\draw[thick] (0, 0) -- (0, \lamopt);

\draw[thick] (\m, 0.1) -- (\m, -0.1) node[below] {$m_1$};
\draw[thick] (0.1, \lamopt) -- (-0.1, \lamopt) node[left] {$\lambda$};

\draw[dashed, gray!60] (0, \lamopt) -- (\m, \lamopt);

\filldraw[fill=blue!10, draw=blue!80!black, thick] (0.5, 1.75) rectangle (7.0, \lamopt);

\filldraw[fill=orange!50, draw=orange!80!black, thick] (0, 0) rectangle (1, 1.4);
\filldraw[fill=orange!50, draw=orange!80!black, thick] (0, 1.4) rectangle (0.5, 2.6);

\draw[|-|, thick] (1, -0.55) -- (4, -0.55) node[midway, below=2pt] {$q$};

\filldraw[fill=orange!20, draw=orange!80!black, thick] (4, 0) rectangle (5, 1.55);
\filldraw[fill=orange!20, draw=orange!80!black, thick] (5, 0) rectangle (6, 1.65);
\filldraw[fill=orange!20, draw=orange!80!black, thick] (6, 0) rectangle (7, 1.75);
\filldraw[fill=orange!20, draw=orange!80!black, thick] (7, 0) rectangle (8, 1.85);

\end{tikzpicture}
\caption{Contiguous.}
\end{subfigure}

\caption{Transformation to reschedule the single job with low height.}
\label{fig:reschedule}
\end{figure}
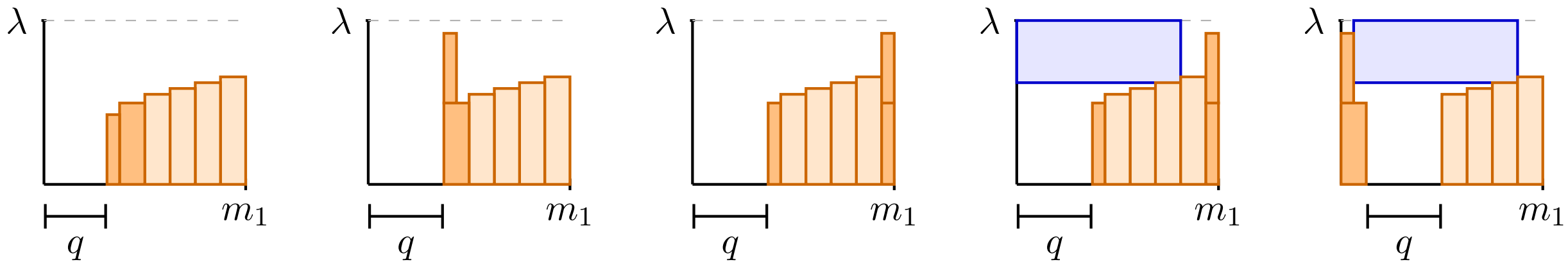

%% file: literatur.bib
@inproceedings{ohnesorge,
  author    = {Klaus Jansen and
               Felix Ohnesorge},
  title     = {A Practical 73/50 Approximation for Contiguous Monotone Moldable Job
               Scheduling},
  booktitle = {43rd International Symposium on Theoretical Aspects of Computer Science,
               {STACS} 2026},
  series    = {LIPIcs},
  publisher = {Schloss Dagstuhl - Leibniz-Zentrum f{\"{u}}r Informatik},
  year      = {2026},
  doi       = {10.4230/LIPICS.STACS.2026.56},
  bibsource = {dblp computer science bibliography, https://dblp.org}
}

@article{DBLP:journals/siamcomp/MounieRT07,
  author    = {Gregory Mounie and
               Christophe Rapine and
               Denis Trystram},
  title     = {A 3/2-Approximation Algorithm for Scheduling Independent Monotonic
               Malleable Tasks},
  journal   = {{SIAM} J. Comput.},
  volume    = {37},
  number    = {2},
  pages     = {401--412},
  year      = {2007},
  doi       = {10.1137/S0097539701385995},
  bibsource = {dblp computer science bibliography, https://dblp.org}
}

@inproceedings{DBLP:conf/ipps/JansenL18,
  author    = {Klaus Jansen and
               Felix Land},
  title     = {Scheduling Monotone Moldable Jobs in Linear Time},
  booktitle = {2018 {IEEE} International Parallel and Distributed Processing Symposium,
               {IPDPS} 2018, Vancouver, BC, Canada, May 21-25, 2018},
  pages     = {172--181},
  publisher = {{IEEE} Computer Society},
  year      = {2018},
  doi       = {10.1109/IPDPS.2018.00027},
  bibsource = {dblp computer science bibliography, https://dblp.org}
}

@inproceedings{DBLP:conf/europar/GrageJO23,
  author    = {Kilian Grage and
               Klaus Jansen and
               Felix Ohnesorge},
  title     = {Improved Algorithms for Monotone Moldable Job Scheduling Using Compression
               and Convolution},
  booktitle = {Euro-Par 2023: Parallel Processing - 29th International Conference
               on Parallel and Distributed Computing},
  series    = {Lecture Notes in Computer Science},
  year      = {2023},
  doi       = {10.1007/978-3-031-39698-4\_34},
  bibsource = {dblp computer science bibliography, https://dblp.org}
}

@article{DBLP:journals/eor/WuZC23,
  author    = {Fangfang Wu and
               Xiandong Zhang and
               Bo Chen},
  title     = {An improved approximation algorithm for scheduling monotonic moldable
               tasks},
  journal   = {Eur. J. Oper. Res.},
  volume    = {306},
  number    = {2},
  pages     = {567--578},
  year      = {2023},
  doi       = {10.1016/J.EJOR.2022.08.034},
  bibsource = {dblp computer science bibliography, https://dblp.org}
}

@article{DBLP:journals/siamdm/DuL89a,
  author    = {Jianzhong Du and
               Joseph Y.{-}T. Leung},
  title     = {Complexity of Scheduling Parallel Task Systems},
  journal   = {{SIAM} J. Discret. Math.},
  volume    = {2},
  number    = {4},
  pages     = {473--487},
  year      = {1989},
  doi       = {10.1137/0402042},
  bibsource = {dblp computer science bibliography, https://dblp.org}
}

@article{DBLP:journals/scheduling/Johannes06,
  author    = {Berit Johannes},
  title     = {Scheduling parallel jobs to minimize the makespan},
  journal   = {J. Sched.},
  volume    = {9},
  number    = {5},
  pages     = {433--452},
  year      = {2006},
  url       = {https://doi.org/10.1007/s10951-006-8497-6},
  doi       = {10.1007/S10951-006-8497-6},
  bibsource = {dblp computer science bibliography, https://dblp.org}
}

@inproceedings{DBLP:conf/icalp/JansenT08,
  author    = {Klaus Jansen and
               Ralf Th{\"{o}}le},
  title     = {Approximation Algorithms for Scheduling Parallel Jobs: Breaking the
               Approximation Ratio of 2},
  booktitle = {Automata, Languages and Programming, 35th International Colloquium,
               {ICALP} 2008},
  series    = {Lecture Notes in Computer Science},
  pages     = {234--245},
  publisher = {Springer},
  year      = {2008},
  doi       = {10.1007/978-3-540-70575-8\_20},
  bibsource = {dblp computer science bibliography, https://dblp.org}
}

@inproceedings{DBLP:conf/esa/JansenR19,
  author    = {Klaus Jansen and
               Malin Rau},
  title     = {Closing the Gap for Pseudo-Polynomial Strip Packing},
  booktitle = {27th Annual European Symposium on Algorithms, {ESA} 2019},
  series    = {LIPIcs},
  pages     = {62:1--62:14},
  publisher = {Schloss Dagstuhl - Leibniz-Zentrum f{\"{u}}r Informatik},
  year      = {2019},
  url       = {https://doi.org/10.4230/LIPIcs.ESA.2019.62},
  doi       = {10.4230/LIPICS.ESA.2019.62},
  bibsource = {dblp computer science bibliography, https://dblp.org}
}

@inproceedings{DBLP:conf/wea/GermannJOT24,
  author    = {Sebastian Bernhard Germann and
               Klaus Jansen and
               Felix Ohnesorge and
               Malte Tutas},
  title     = {3/2-Dual Approximation for {CPU/GPU} Scheduling},
  booktitle = {22nd International Symposium on Experimental Algorithms, {SEA} 2024},
  year      = {2024},
  doi       = {10.4230/LIPICS.SEA.2024.13},
  bibsource = {dblp computer science bibliography, https://dblp.org}
}

@article{DBLP:journals/jacm/HochbaumS87,
  author    = {Dorit S. Hochbaum and
               David B. Shmoys},
  title     = {Using dual approximation algorithms for scheduling problems theoretical
               and practical results},
  journal   = {J. {ACM}},
  volume    = {34},
  number    = {1},
  pages     = {144--162},
  year      = {1987},
  doi       = {10.1145/7531.7535},
  bibsource = {dblp computer science bibliography, https://dblp.org}
}

@article{DBLP:journals/siamam/Graham69,
  author    = {Ronald L. Graham},
  title     = {Bounds on Multiprocessing Timing Anomalies},
  journal   = {{SIAM} Journal of Applied Mathematics},
  volume    = {17},
  number    = {2},
  pages     = {416--429},
  year      = {1969},
  bibsource = {dblp computer science bibliography, https://dblp.org}
}

@article{DBLP:journals/siamdm/Jansen10,
  author    = {Klaus Jansen},
  title     = {An {EPTAS} for Scheduling Jobs on Uniform Processors: Using an {MILP}
               Relaxation with a Constant Number of Integral Variables},
  journal   = {{SIAM} J. Discret. Math.},
  volume    = {24},
  number    = {2},
  pages     = {457--485},
  year      = {2010},
  url       = {https://doi.org/10.1137/090749451},
  doi       = {10.1137/090749451},
  bibsource = {dblp computer science bibliography, https://dblp.org}
}

@article{DBLP:journals/siamcomp/CoffmanGJ78,
  author    = {Edward G. Coffman Jr. and
               M. R. Garey and
               David S. Johnson},
  title     = {An Application of Bin-Packing to Multiprocessor Scheduling},
  journal   = {{SIAM} J. Comput.},
  volume    = {7},
  number    = {1},
  pages     = {1--17},
  year      = {1978},
  doi       = {10.1137/0207001},
  bibsource = {dblp computer science bibliography, https://dblp.org}
}

@article{DBLP:journals/concurrency/BleuseKMMT15,
  author    = {Rapha{\"{e}}l Bleuse and
               Safia Kedad{-}Sidhoum and
               Florence Monna and
               Gr{\'{e}}gory Mouni{\'{e}} and
               Denis Trystram},
  title     = {Scheduling independent tasks on multi-cores with {GPU} accelerators},
  journal   = {Concurr. Comput. Pract. Exp.},
  volume    = {27},
  number    = {6},
  pages     = {1625--1638},
  year      = {2015},
  doi       = {10.1002/CPE.3359},
  bibsource = {dblp computer science bibliography, https://dblp.org}
}

@article{drozdowski1995no32approx,
  title             = {On the complexity of multiprocessor task scheduling},
  author            = {Maciej Drozdowski},
  journal           = {Bulletin of The Polish Academy of Sciences-technical Sciences},
  year              = {1995},
  volume            = {43},
  pages             = {381-392},
  _bib2doi_finished = {true}
}

@article{land2013boundingmoldable,
  author             = {Klaus Jansen and Felix Land and Kati Land},
  title              = {Bounding the Running Time of Algorithms for Scheduling and Packing Problems},
  journal            = {{SIAM} J. Discret. Math.},
  volume             = {30},
  number             = {1},
  pages              = {343--366},
  year               = {2016},
  doi                = {10.1137/140952636},
  timestamp          = {Sat, 25 Apr 2020 01:00:00 +0200},
  biburl             = {https://dblp.org/rec/journals/siamdm/JansenLL16.bib},
  bibsource          = {dblp computer science bibliography, https://dblp.org},
  _bib2doi_selected  = {dblp:/rec/journals/siamdm/JansenLL16.bib},
  _bib2doi_confirmed = {true}
}

@inproceedings{jansen2002linear,
  title              = {Linear-Time Approximation Schemes for Scheduling Malleable Parallel Tasks},
  author             = {Klaus Jansen and Lorant Porkolab},
  booktitle          = {Proceedings of the tenth Annual ACM-SIAM Symposium on Discrete Algorithms (SODA 1999)},
  pages              = {490--498},
  year               = {1999},
  publisher          = {{ACM/SIAM}},
  timestamp          = {Fri, 07 Dec 2012 00:00:00 +0100},
  biburl             = {https://dblp.org/rec/conf/soda/JansenP99.bib},
  bibsource          = {dblp computer science bibliography, https://dblp.org},
  url                = {http://dl.acm.org/citation.cfm?id=314500.314870},
  _bib2doi_selected  = {dblp:/rec/conf/soda/JansenP99.bib},
  _bib2doi_confirmed = {true}
}

@article{decker200654,
  title              = {A 54-approximation algorithm for scheduling identical malleable tasks},
  author             = {Decker, Thomas and L{\"u}cking, Thomas and Monien, Burkhard},
  journal            = {Theoretical Computer Science},
  volume             = {361},
  number             = {2-3},
  pages              = {226--240},
  year               = {2006},
  publisher          = {Elsevier},
  timestamp          = {Tue, 26 Sep 2023 01:00:00 +0200},
  bibsource          = {dblp computer science bibliography, https://dblp.org},
  doi                = {10.1016/j.tcs.2006.05.012},
  _bib2doi_selected  = {dblp:/rec/journals/tcs/DeckerLM06.bib},
  _bib2doi_confirmed = {true}
}

@inproceedings{turek1992approximate,
  title              = {Approximate algorithms scheduling parallelizable tasks},
  author             = {Turek, John and Wolf, Joel L and Yu, Philip S},
  booktitle          = {Proceedings of the fourth annual ACM symposium on Parallel algorithms and architectures (SPAA 1992)},
  pages              = {323--332},
  year               = {1992},
  timestamp          = {Wed, 21 Nov 2018 00:00:00 +0100},
  biburl             = {https://dblp.org/rec/conf/spaa/TurekWY92.bib},
  bibsource          = {dblp computer science bibliography, https://dblp.org},
  doi                = {10.1145/140901.141909},
  _bib2doi_selected  = {dblp:/rec/conf/spaa/TurekWY92.bib},
  _bib2doi_confirmed = {true}
}

@inproceedings{ludwig1994,
  author             = {Walter Ludwig and Prasoon Tiwari},
  title              = {Scheduling Malleable and Nonmalleable Parallel Tasks},
  booktitle          = {Proceedings of the Fifth Annual {ACM-SIAM} Symposium on Discrete Algorithms, (SODA 1994)},
  pages              = {167--176},
  publisher          = {{ACM/SIAM}},
  year               = {1994},
  url                = {http://dl.acm.org/citation.cfm?id=314464.314491},
  timestamp          = {Fri, 07 Dec 2012 00:00:00 +0100},
  biburl             = {https://dblp.org/rec/conf/soda/LudwigT94.bib},
  bibsource          = {dblp computer science bibliography, https://dblp.org},
  _bib2doi_selected  = {dblp:/rec/conf/soda/LudwigT94.bib},
  _bib2doi_confirmed = {true}
}

@book{DBLP:books/kellerer/knapsackproblems,
  author    = {Hans Kellerer and
               Ulrich Pferschy and
               David Pisinger},
  title     = {Knapsack problems},
  publisher = {Springer},
  year      = {2004},
  doi       = {10.1007/978-3-540-24777-7},
  isbn      = {978-3-540-40286-2},
  bibsource = {dblp computer science bibliography, https://dblp.org}
}
